\documentclass[a4paper,fleqn,usenatbib]{mnras}

\usepackage{graphicx,psfrag}
\usepackage{mathrsfs}
\usepackage{amsmath,amsfonts,amssymb}
\usepackage{multirow}
\usepackage{comment}
\usepackage{units}
\usepackage{enumerate}
\usepackage[normalem]{ulem}

\newcommand{\be}{\begin{equation}}
\newcommand{\ee}{\end{equation}}
\newcommand{\bea}{\begin{eqnarray}}
\newcommand{\eea}{\end{eqnarray}}
\newcommand{\bel}{\begin{align}}
\newcommand{\eel}{\end{align}}

\def\GMc2{{\rm G M_{\odot} c^{-2}}}

\usepackage{color}

\definecolor{cyan}{rgb}{0,0.9,0.9}
\definecolor{orange}{rgb}{0.9,0.5,0}
\definecolor{magenta}{rgb}{1,0,1}
\definecolor{purple}{rgb}{0.8,0.4,0.8}
\definecolor{gray}{rgb}{0.8242,0.8242,0.8242}

\title{Disk formation in the collapse of supramassive neutron stars}

\author[G.~Camelio, T.~Dietrich, and S.~Rosswog]{Giovanni Camelio$^1$,
Tim Dietrich$^{2,3}$,
Stephan Rosswog$^1$\\
${}^1$ Astronomy and Oskar Klein Centre, Stockholm University, AlbaNova, SE-10691, Stockholm, Sweden\\
${}^2$ Nikhef, Science Park, 1098 XG Amsterdam, The Netherlands \\
${}^3$ Max Planck Institute for Gravitational Physics (Albert Einstein Institute), Am M\"uhlenberg 1, Potsdam 14476, Germany}

\date{\today}
\pubyear{2018}

\begin{document}
\label{firstpage}
\pagerange{\pageref{firstpage}--\pageref{lastpage}}
\maketitle

\begin{abstract}
Short gamma-ray bursts (sGRBs) show a large diversity in their properties.
This suggests that the observed phenomenon can be caused by different
``central engines'' or that the engine produces a variety of outcomes
depending on its parameters, or possibly both. The most popular
engine scenario, the merger of two neutron stars, has received support
from the recent Fermi and INTEGRAL detection of a burst of gamma rays (GRB170817A)
following the neutron star merger GW170817, but at the moment it is
not clear how peculiar this event potentially was.
Several sGRBs engine models involve the collapse of a supramassive
neutron star that produces
a black hole plus an accretion disk.
We study this scenario for a variety of equations of states
both via angular momentum considerations based on equilibrium models
and via fully dynamical Numerical Relativity simulations. We obtain
a broader range of disk forming configurations than earlier studies
but we agree with the latter that none of these
configurations is likely to produce a phenomenon that would be
classified as an sGRB.
\end{abstract}


\section{Introduction}
\label{sec:intro}
With the first detection of a neutron star merger in both gravitational
\citep{abbott17b} and electromagnetic waves \citep{abbott17c, Coulter+2017} the era of
multi-messenger astrophysics has begun in earnest. This single event
brought a major leap forward for a number of areas: it allowed for a new,
independent measurement of the Hubble constant \citep{abbott17a}, it
conclusively established that neutron star mergers are a major cosmic
source of r-process elements 
\citep{lattimer74,Eichler:1989ve,rosswog99,freiburghaus99b,cowperthwaite17,Smartt:2017fuw,kasliwal17,Kasen:2017sxr,Tanvir:2017pws,Rosswog:2017sdn},
and, with precise limits on the propagation speed of gravitational waves
\citep{abbott17c}, it placed strict constraints on alternative theories
of gravity. Moreover, the triggering of the Fermi and INTEGRAL satellites on a
short gamma-ray burst (sGRB) 1.7 seconds after the gravitational wave
(GW) peak lends support to the long-held conjecture that neutron star
mergers produce GRBs \citep{Paczynski:1986px,Eichler:1989ve}. It has, however,
been debated whether this GRB event was an intrinsically sub-luminous one
with $E_{\gamma,{\rm iso}}\sim 6 \times 10^{46}$ erg
\citep{kasliwal17,mooley17,nakar18} or a typical short
GRB with $\sim 10^{50}-10^{52}$ erg \citep{Berger:2013jza,fong15}, but
seen off axis, see for example \cite{margutti18,Lyman2018}.\\
In general, sGRBs exhibit  a large variety of properties and it is
not well understood how this diversity relates to the central engine(s). 
A particularly puzzling property is late-time X-ray activity
on time scales that exceed the dynamical time scales of a compact engine
($\sim 1$ ms) by many orders of magnitude, see e.g.
\cite{villasenor05,barthelmy05,Rowlinson+2013,Gompertz+2014}. 
One possibility would be that the sGRB is produced by a magnetar \citep{Metzger2011, Bucciantini2012},
provided that excessive ``baryonic pollution'' e.g.{} due to a neutrino-driven wind \citep{Dessart2009, Perego2014}
can be avoided, otherwise the outflow will be choked \citep{Murguia2017}.
Also models involving quark stars have been suggested \citep{Drago2016, Pili2016}.
Alternatively, \citet{macfadyen05} proposed that
such bursts could be caused by neutron stars (NSs) accreting from a non-degenerate
companion star. Upon collapse, a black hole (BH) plus accretion disk system 
would form and launch the relativistic outflow that produces the GRB.
The late X-ray activity would result from the interaction of the outflow with the extended
companion star.
In a black hole accretion flow, a fraction of the accreted rest mass energy
is released as radiation \citep{Frank2002book}.
Therefore, to produce 
$E_{\gamma,\rm iso}$ as GRB energy, the accretion disk would need to have a mass of the order of
\begin{equation}
M_{\rm disk} \sim 2 \times 10^{-4} {\rm M_\odot} \left(\frac{E_{\rm iso}}{10^{51} \; \rm erg} \right)
\left(\frac{f_b}{1/50}\right) \left(\frac{0.05}{\varepsilon}\right),\\
\label{eq:mdisk}
\end{equation}
where $M_{\rm disk}$ is the disk mass, $\varepsilon$ is the accretion efficiency and
$f_b= \Delta\Omega/4\pi$ is the beaming fraction.
The late X-ray activity is also addressed in so-called 
``time-reversal scenarios'' \citep{Ciolfi+Siegel2015,Rezzolla:2014nva}
where a long-lived supramassive neutron star produces the long-lasting X-ray 
emission which initially is trapped in an optically thick nebula. As in the
scenario proposed by \cite{macfadyen05}, also here it is crucial for the
model that at some point the supramassive neutron star collapses to 
black hole plus torus system to launch the GRB.\\
The question whether such a collapse really produces an accretion torus that 
is massive enough for launching a typical sGRB, has recently been addressed
by  \cite{Margalit:2015qza}. They constructed
rapidly rotating neutron stars using the RNS code \citep{Stergioulas:1994ea}
and studied the corresponding angular momentum distribution. The authors came to the conclusion
that it is unlikely that an accretion disk massive enough to launch an energetic 
GRB can be formed. In this paper, we revisit this problem. We construct our initial conditions
with the XNS code \citep{Bucciantini+Del_Zanna2011, Pili+2014},
that makes use of the extended conformal flatness approximation
of \cite{Cordero-Carrion+2009} and we study the angular momentum spectrum
to estimate the resulting disk mass after collapse. We scrutinize our conclusions
by simulating for a selected set of configurations the collapse directly with 
fully dynamical Numerical Relativity simulations. We summarize our numerical methods
in Sec.~\ref{sec:methods}, discuss the rotating equilibrium configurations in Sec.~\ref{sec:equilibrium}
and describe the dynamical collapse simulations in Sec.~\ref{sec:evolution}. 
Our results are summarized in Sec.~\ref{sec:conclusions}.
Comparisons between XNS and RNS results are provided in Appendix~\ref{app:XNS_vs_RNS}.

\section{Numerical Methods}
\label{sec:methods}

\subsection{Governing Equations}
\label{ssec:eos}

\begin{figure}
\includegraphics[width=\columnwidth]{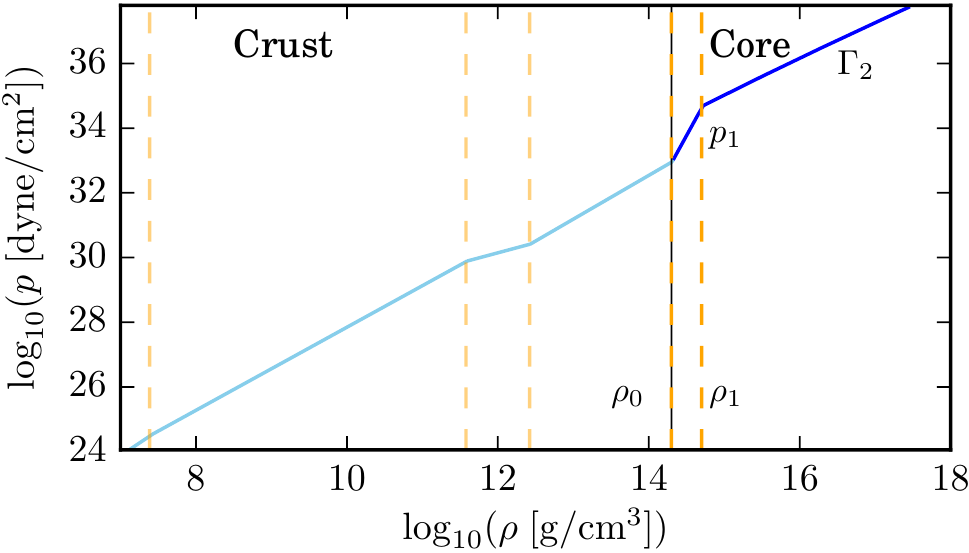}
\caption{Example of the six-piece polytropic EOS employed in this work. 
         The light blue line refers to the SLy crust employed at low densities,
         while the dark blue line refers to the high density part of the EOS. We
         mark the transition between the different polytropic pieces with dashed orange
         lines.  The thin black vertical line corresponds to $\rho_0 = \unit[10^{14.3}]{g/cm^3}$, namely
         the density at which the crust is attached to the high density EOS.}
         \label{fig:eos}
\end{figure}

Our goal is to construct rigidly rotating neutron stars as pre-collapse initial
conditions. We assume stationarity and axisymmetry and therefore can
write the metric in quasi-isotropic coordinates 
($t,r,\theta,\phi$)~\citep{Gourgoulhon:2011} as
\begin{multline}
\label{eq:metric}
\mathrm ds^2 = -N^2\mathrm dt^2 + A^2(\mathrm dr^2 + r^2\mathrm d\theta^2)\\
+r^2B^2\sin^2\theta(\mathrm d\phi -\omega\mathrm dt)^2.
\end{multline}
Here the cylindrical radius is defined as $R= B r \sin(\theta)$. 
The metric functions $N,A,B,\omega$ depend purely on $r$ and $\theta$, where
$N$ denotes the lapse and $\omega$ is the intrinsic angular velocity of the
zero angular momentum observer (ZAMO) relative to infinity.\footnote{$\omega$ is non-zero because of
the frame dragging effect due to the rotation of the neutron star.}
The first integral \citep{Gourgoulhon:2011}
for a cold equation of state (EOS) in rigid rotation is
\begin{align}
\label{eq:1st_integral}
\ln h + \ln N - \ln W ={}& \mathrm{const},\\
h={}& \frac{\epsilon+p}\rho,\\
W={}&\left(1 -U^2\right)^{-1/2},
\end{align}
where $h$ is the specific enthalpy, $\epsilon$ the total energy density, $p$ the pressure,
$\rho$ the
rest mass density, $W$ the Lorentz factor, and $U$ the magnitude of the fluid 3-velocity
in the ZAMO reference frame. The fluid 3-velocity can be determined from the
relations
\begin{equation}
U={}R\ |U^\phi|, \quad {\rm and} \quad U^\phi={}\frac{\Omega-\omega}{N},
\end{equation}
where $U^\phi$ is the $\phi$-component of the contravariant fluid 3-velocity
and $\Omega$ is the star's angular velocity as seen by an observer at infinity,
$\Omega\equiv \Omega(r,\theta)$.

Since the system we consider in this study can
be considered to a very good approximation as being in cold
$\beta$-equilibrium, we restrict ourselves to a barotropic EOS for which
temperature effects can be neglected and all particle species are in
equilibrium. Consequently, the EOS solely depends 
on the density, $p=p(\rho)$.
A common choice is to impose a polytropic EOS of the type 
\begin{equation}
\label{eq:p_rho}
p(\rho)=K\rho^\Gamma,
\end{equation}
where $K$ is called polytropic constant and $\Gamma$ polytropic exponent.
We combine  multiple polytropes in different density regions
to approximate more complicated and realistic EOSs (see e.g.~\citealp{Read:2008iy}).
The obtained piecewise polytropic EOSs are still barotropic and 
hold as long as temperature effects can be neglected. 
Within this article we follow \citet{Read:2008iy} and \citet{Margalit:2015qza}
in the construction of piecewise polytropic EOSs and assume a 
low density part ($\rho<\rho_0=\unit[10^{14.3}]{\rm g\,cm^{-3}}$) 
approximating an SLy crust \citep{Douchin:2001sv,Read:2008iy}, 
see Tab.~\ref{tab:eos} and the light blue line in Fig.~\ref{fig:eos}.
The high density part ($\rho>\rho_0=\unit[10^{14.3}]{\rm g\,cm^{-3}}$) is a
2-pieces polytrope;
see Tab.~\ref{tab:eos} and the dark blue line in Fig.~\ref{fig:eos}. 
In this high density part, the first polytrope ($\rho_0<\rho<\rho_1=\unit[10^{14.7}]{\rm g\,cm^{-3}}$)
is determined on one side by the SLy EOS that fixes the value of the pressure
$p(\rho_0)$ and on the other side by the free parameter $p_1\equiv p(\rho_1)$.
The second polytrope ($\rho>\rho_1$) is determined by $p_1$ and
by the polytropic exponent $\Gamma_2$. Therefore, the whole EOS is
fixed by only two parameters, $p_1$ and $\Gamma_2$, cf.~Tab.~\ref{tab:eos}.
\begin{table}
  \centering    
  \caption{Parameters of the piecewise polytropic EOS adopted in this paper.
  Density range $\rho$, polytropic exponent $\Gamma$, and polytropic constant $K$
  are reported in each column.
  The horizontal line separates the low-density SLy crust from the high-density
  core EOS. The quantities denoted with an asterisk `$\ast$' depend on the choice
  of $p_1$ and $\Gamma_2$. Beware that the dimensions of $K$ depend on $\Gamma$.}
  \begin{tabular}{ccc}
    \hline
    $\rho\ [\unit{g\,cm^{-3}}]$  & $\qquad\Gamma\qquad$ & $K/c^2\ [\unit{g\,cm^{-3}}]^{1-\Gamma}$ \\ 
    \hline
    $<2.44034\times10^7$ & $1.58425$ & $6.80110\times10^{-9}$ \\
    $<3.78358\times10^{11}$ & $1.28733$ & $1.06186\times10^{-6}$ \\
    $<2.62780\times10^{12}$ & $0.62223$ & $5.32697\times 10^{+1} $    \\
    $<\rho_0\equiv1.99526\times10^{14}$ & $1.35692$ & $3.99874\times10^{-8}$ \\
    \hline
    $<\rho_1\equiv5.01187\times10^{14}$ & $\ast$ & $\ast$ \\
    $>\rho_1\equiv5.01187\times10^{14}$ & $\Gamma_2$ & $\ast$\\
    \hline
  \end{tabular}
  \label{tab:eos}
\end{table}

In the computation of the equilibrium configurations we 
neglect thermal effects, but (apart from one test) we include them 
for the dynamical simulations, see Sec.~\ref{sec:evolution}, 
by adding a thermal pressure component
to the barotropic pressure $p(\rho)$
\begin{equation}
 p(\rho, \epsilon) = p(\rho) +  \epsilon \rho (\Gamma_{\rm th}-1).\label{eq:p_evolution}
\end{equation}
In accordance with previous work and following the discussion in \citet{Bauswein:2010dn}, 
where full tabulated EOSs are compared against the approximate description of 
Eq.~\eqref{eq:p_evolution}, we employ $\Gamma_{\rm th}=1.75$. 
Additionally, we also perform a dynamical simulation without 
additional thermal component to allow for an assessment of systematic uncertainties.

\subsection{Equilibrium configurations}
\label{ssec:XNS}

Our investigation of the equilibrium configurations for different EOSs 
is based on the XNS code \citep{Bucciantini+Del_Zanna2011, Pili+2014}, which determines
the rotating
stellar configuration in quasi-isotropic coordinates under the extended conformal
flatness approximation \citep{Cordero-Carrion+2009}. In the extended conformal
flatness approximation, all elliptic equations that characterize the spacetime 
metric are hierarchically decoupled, which leads to a simplified metric with
\begin{equation} 
A(r,\theta)\equiv B(r,\theta)\equiv\psi^2(r,\theta),
\end{equation} 
where $\psi$ denotes the conformal factor. 
The approximation is justified since the metric functions 
$A$ and $B$ typically differ at most by
about $0.1\%$ \citep{Gourgoulhon:2011}. \\
The XNS code originally descends from the X-ECHO code \citep{Bucciantini+Del_Zanna2011}
and therefore inherits some features not needed for our purposes.
Its main focus is on the interplay between rigid or differential rotation
and poloidal and/or magnetic fields \citep{Pili2017}.
For this study we have modified the publicly available version
of XNS to the following workflow: 
\begin{enumerate}[(i)]
\item set the target stellar parameters central rest mass density,  $\rho_c$,
and angular speed seen by an observer at infinity,
$\Omega$.
\item determine the initial configuration from the TOV solution (Tolman-Oppenheimer-Volkoff,
namely the one that describe a spherical neutron star) with central density $\rho_c$,
or if available, load a previously relaxed configuration from a sequence,
obtained for example in the search for the Keplerian configuration.
\item repeat until
$\max_{r,\theta}|\rho_\mathrm{old}(r,\theta)-\rho_\mathrm{new}(r,\theta)|<
   \unit[10^{-9}]{M_\odot^{-2}}$ (in units $c=G=1$):
   \begin{enumerate}[(a)]
      \item using the old metric and matter quantities, solve the hierarchically
      decoupled equations of the extended conformal flatness approximation
      and update the metric fields.
      \item update the matter fields solving the first integral, Eq.~\eqref{eq:1st_integral}, with
      central density $\rho_c$ and angular velocity $\Omega$.
   \end{enumerate}
\end{enumerate}

The main differences between our workflow and the original 
one are that we update the matter fields only through the first integral inversion avoiding
conservative-to-primitive variable inversion. We directly set the central density
in the first integral instead of using an external root-finding
cycle, and we allow for an initial configuration other than the TOV one.
Additional major technical modifications are the adoption of an inner (uniformly spaced) and an outer
(increasingly spaced) radial grid, an angular grid defined on the
Gauss-Legendre quadrature points, and the use of a true vacuum outside
the neutron star instead of an artificial atmosphere.

Our modified XNS version is ten times
faster than RNS\footnote{On a 1.40GHz CPU (Intel(R) Core(TM) i3-2365M) and
4GB RAM laptop with \texttt{-O2}
optimization.}. This is in part due to the hierarchical decoupling of
the equations for the spacetime metric in the extended conformal flatness approximation.
In Appendix~\ref{app:XNS_vs_RNS} we present a detailed convergence
study and compare the results of our modified  XNS version with the 
publicly available RNS code \citep{Stergioulas:1994ea}.
In particular, we find that XNS
recovers the stellar properties within the precision of the extended conformal
flatness approximation and yield practically identical results of RNS. 
For the exploration of  the parameters space we employ XNS, while the initial configurations
which we evolve dynamically with the BAM code are constructed with RNS, since the 
interface between RNS and BAM has been implemented and tested in detail in 
a previous work \citep{Dietrich:2014wja}.

\subsection{Dynamical evolution}
\label{ssec:BAM}

For the dynamical evolution we solve 
Einstein's field equations in their 3+1 form 
recast in the Z4c evolution system \citep{Bernuzzi:2009ex,Hilditch:2012fp}. 
The gauge sector employs the 1+log and 
gamma-driver equations developed for black holes in the moving
puncture approach \citep{Bona:1994a,Alcubierre:2002kk,vanMeter:2006vi,
Campanelli:2005dd,Baker:2005vv}.
This particular gauge choice, often called `puncture
gauge', handles automatically the gravitational 
collapse of a neutron star to a black hole as discussed 
in \citet{Baiotti:2007np,Thierfelder:2010dv,Dietrich:2014wja} and is 
therefore  particularly well-suited for our study. 

The simulations are performed with the BAM 
code \citep{Brugmann:2008zz,Thierfelder:2011yi}. 
BAM employs the method of lines approximating spatial derivatives of the metric variables
by 4th order finite differences. Time integration is performed with an explicit 
4th order Runge-Kutta scheme. 
The grid used in this work consists of a hierarchy of $7$ cell-centered nested Cartesian 
boxes. Every box $l=0,...,6$ employs a constant grid spacing $h_l$ and 
$n$ points per direction. Boxes use a $2:1$ refinement strategy, i.e., 
each coarser box employs a grid spacing $h_{l-1}=2 h_l$.
For the time stepping of the mesh refinement, we employ
the Berger-Oliger algorithm \citep{Berger:1984zza} extended by a refluxing step that enforces 
energy and momentum conservation across refinement 
boundaries \citep{Berger:1989a,East:2011aa,Reisswig:2012nc,Dietrich:2015iva}. 
The equations of GRHD are solved with
a standard high-resolution-shock-capturing (HRSC) scheme based
on primitive reconstruction and the Local-Lax-Friedrich
central scheme for the numerical fluxes. The primitive 
reconstruction uses a fifth-order Weighted Essentially
Non-Oscillatory (WENO) scheme \citep{Borges:2008a,Bernuzzi:2012ci}, 
called WENOZ. 
Other limiters are used for comparison to
assess the numerical uncertainties, see Tab.~\ref{tab:resolution}. 
The simulations presented in this article employ quadrant
symmetry to reduce computational costs. 

An important detail that is of particular 
interest for our study is the artificial atmosphere which
is needed by GRHD simulations. 
As described in \citet{Thierfelder:2011yi} and \citet{Dietrich:2015iva}, 
we use a low-density static and barotropic atmosphere at a density level
\begin{equation}
\rho_{\rm atm} = f_{\rm atm}\cdot \text{max}[\rho(t=0)].
\label{eq:atm}
\end{equation}
During the inversion from conservative to primitive variables
we set a grid point to the atmosphere values if the density falls below the
threshold 
\begin{equation}
\rho_{\rm thr} = f_{\rm thr}\cdot \rho_{atm}. 
\end{equation}
Throughout this work we employ for the threshold $f_{\rm thr}=10^1$ and
vary the value $f_{\rm atm}=10^{-18},10^{-19},10^{-20}$ to understand the effects
of the artificial atmosphere on the debris disk mass. 
We also study numerical uncertainties by employing different grid resolutions. 
All employed combinations of resolutions, flux limiters, 
and atmosphere values are summarized in Tab.~\ref{tab:resolution}.

\begin{table}
\renewcommand{\arraystretch}{1.2}
  \centering    
  \caption{Configurations employed for the dynamical evolutions. 
  The columns refer to: configuration name, number of points in the 
  Cartesian boxes, grid spacing in the refinement 
  level covering the neutron star, atmosphere factor [see Eq.~\eqref{eq:atm}], and flux limiter: 
  WENOZ \protect\citep{Borges:2008a,Bernuzzi:2012ci}, linear total variation diminishing (LINTVD) \protect\citep{Shu:1989}, 
  3rd order Essentially-Non-Oscillatory 3rd-order method (CENO3) \protect\citep{Liu:1998,DelZanna:2002rv}.
  In addition to the listed setups, we also employ for one physical configuration 
  the setup ${\rm Res2_{atm19}^{WENOZ}}$ but with zero thermal 
  contribution, Eq.~\eqref{eq:p_evolution}. 
  This setup is labeled as ${\rm Res2_{atm19cold}^{WENOZ}}$.}
  \begin{tabular}{l|cccc}        
    \hline
    Name           & $n$    & $h_6 \ [M_\odot]$   & $f_{\rm atm}$ & Limiter\\ 
    \hline
    ${\rm Res1_{atm19}^{WENOZ}}$  & $120$  & $0.1250$ & $10^{-19}$ & ${\rm WENOZ}$  \\    
    ${\rm Res2_{atm19}^{WENOZ}}$  & $180$  & $0.0833$ & $10^{-19}$ & ${\rm WENOZ}$  \\
    ${\rm Res3_{atm19}^{WENOZ}}$  & $240$  & $0.0625$ & $10^{-19}$ & ${\rm WENOZ}$  \\
    ${\rm Res4_{atm19}^{WENOZ}}$  & $360$  & $0.0417$ & $10^{-19}$ & ${\rm WENOZ}$  \\    
    \hline
    ${\rm Res2_{atm18}^{WENOZ}}$  & $180$  & $0.0833$ & $10^{-18}$ & ${\rm WENOZ}$  \\
    ${\rm Res2_{atm20}^{WENOZ}}$  & $180$  & $0.0833$ & $10^{-20}$ & ${\rm WENOZ}$  \\
    \hline
    ${\rm Res2_{atm18}^{LINTVD}}$  & $180$  & $0.0833$ & $10^{-18}$ & ${\rm LINTVD}$  \\
    ${\rm Res2_{atm20}^{CENO3}}$  & $180$  & $0.0833$ & $10^{-20}$ & ${\rm CENO3}$  \\
    \hline
  \end{tabular}
 \label{tab:resolution}
\end{table}

The initial conditions obtained (i.e., the equilibrium configurations) are ported onto 
the BAM grid by Lagrangian interpolation.

\section{Equilibrium Configurations}
\label{sec:equilibrium}

\begin{figure}
\includegraphics[width=\columnwidth]{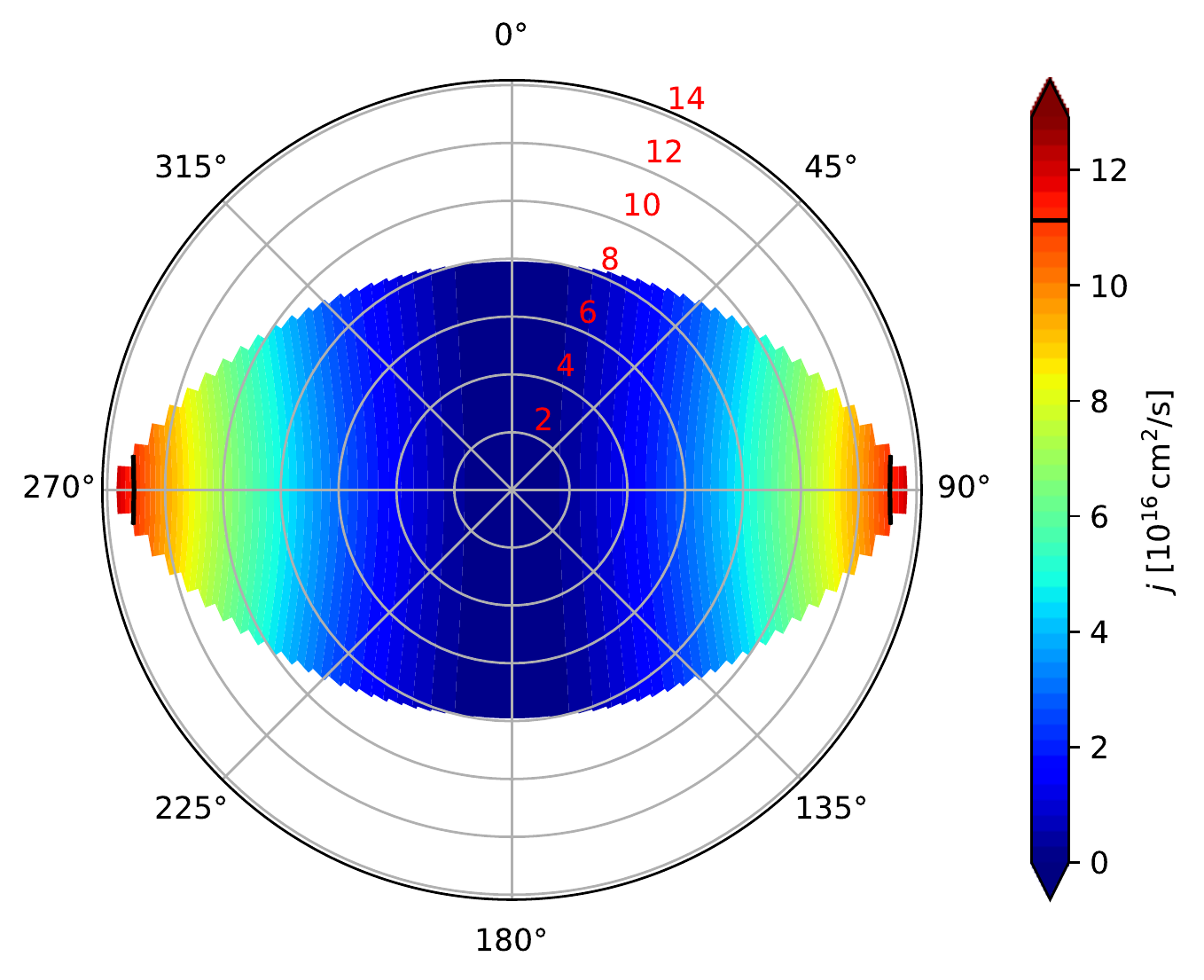}
\caption{Specific angular momentum distribution for a neutron star
in the Case~A configuration
(see text for details). The vertical direction correspond to the rotation axis
and the red labels mark the radial coordinate (in km).
The black solid contour corresponds to the
angular momentum of a particle at the ISCO of a black hole with
the same total mass and angular momentum of the neutron star.
Only the matter on the outer side (dark red area) of the black contour can in principle escape
black hole formation.}
\label{fig:theta_r_j}
\end{figure}

In an axisymmetric dynamical system the spectrum of angular momentum,
i.e., the integrated baryon rest mass of all fluid elements with a specific angular momentum,
is strictly conserved in the absence of viscosity \citep{Stark+Piran:1987}. Even if some
viscosity is present in either Nature or a numerical simulation, the 
(dynamical) collapse timescales are too short for viscosity
effects to become important. This suggests to use 
as a necessary condition for the formation of a debris disk that the
specific angular momentum of a matter element at the stellar equator 
before the collapse is greater than the specific angular momentum of the 
innermost stable circular orbit (ISCO) of a Kerr BH with the same mass $M$ 
and angular momentum $J$ of the progenitor neutron star, see Fig.~\ref{fig:theta_r_j}. This
is also the criterion that has been applied in the study of 
\citet{Margalit:2015qza}.
Since the specific angular momentum increases with the rotational frequency of
the star, one expects that for a given EOS and central density $\rho_c$ the collapse of
a maximally rotating star (that rotates at the Keplerian frequency) will
produce the largest debris disk mass. Furthermore, the configuration should be
unstable to collapse to a BH.

In this section we state the stability and disk formation criteria,
delineate our procedure to find the maximally rotating configurations,
describe our results comparing them with \citet{Margalit:2015qza},
choose the configurations that we further analyze with dynamical simulations
in the next section, and compute possible debris disk masses for a set of 
realistic EOSs constructed in~\cite{Read:2008iy}.

\subsection{Stability condition}
\label{ssec:stability}

If the star is non-rotating, the marginally stable configuration 
obeys (Sec.~10.11 of \citealp{Zeldovich+Novikov1971})
\begin{equation}
\frac{\mathrm dM}{\mathrm d\rho_c}=0.
\end{equation}
Increasing the central density beyond this point the star becomes unstable 
and collapses to a BH. In the non-rotating case, this configuration also 
has the maximal gravitational mass.

For rotating stars the marginally stable criterion has to be modified as
follows \citep{Friedman:1988}
\begin{equation}
\label{eq:stability}
\left.\frac{\partial M}{\partial \rho_c}\right|_J = 0,
\end{equation}
where $J$ is the total angular momentum; this condition is sufficient for instability
\citep{Takami+2011}.
However, $J$ is difficult to access during our computation with the XNS 
code since it uses $\rho_c$ and $\Omega$ as input variables.
One can circumvent this problem with a root finding cycle on $J$.
Another possibility is to rewrite the stability condition~\eqref{eq:stability} as 
\begin{equation}
\label{eq:dMdrho_J}
\left.\frac{\partial M}{\partial \rho_c}\right|_J =
\left.\frac{\partial M}{\partial \rho_c}\right|_\Omega
- \left.\frac{\partial J}{\partial \rho_c}\right|_\Omega \cdot
\left.\frac{\partial M}{\partial \Omega}\right|_{\rho_c} \cdot
\left(\left.\frac{\partial J}{\partial \Omega}\right|_{\rho_c}\right)^{-1}.
\end{equation}
We have obtained Eq.~\eqref{eq:dMdrho_J} from
\begin{align}
\label{eq:dM1}
\mathrm dM(\rho_c,J)={}&
\left.\frac{\partial M}{\partial \rho_c}\right|_J \mathrm d\rho_c
+\left.\frac{\partial M}{\partial J}\right|_{\rho_c} \mathrm dJ,\\
\label{eq:dM2}
\mathrm dM(\rho_c,\Omega)={}&
\left.\frac{\partial M}{\partial \rho_c}\right|_\Omega \mathrm d\rho_c
+\left.\frac{\partial M}{\partial \Omega}\right|_{\rho_c} \mathrm d\Omega,\\
\label{eq:dJ}
\mathrm dJ(\rho_c,\Omega)={}&
\left.\frac{\partial J}{\partial \rho_c}\right|_\Omega \mathrm d\rho_c
+\left.\frac{\partial J}{\partial \Omega}\right|_{\rho_c} \mathrm d\Omega.
\end{align}
One first equates Eqs.~\eqref{eq:dM1} and~\eqref{eq:dM2} (the total variation
of the mass should be the same no matter which are the independent variables),
then substitutes Eq.~\eqref{eq:dJ} and equates the terms that
multiply $\mathrm d\rho_c$. Eq.~\eqref{eq:dMdrho_J} is finally obtained using
\begin{equation}
\label{eq:dMdJ_rho}
\left.\frac{\partial M}{\partial J}\right|_{\rho_c} =
\left.\frac{\partial M}{\partial \Omega}\right|_{\rho_c} \cdot
\left.\frac{\partial \Omega}{\partial J}\right|_{\rho_c} =
\left.\frac{\partial M}{\partial \Omega}\right|_{\rho_c} \cdot
\left(\left.\frac{\partial J}{\partial \Omega}\right|_{\rho_c}\right)^{-1},
\end{equation}
where the equalities follow from the fact that, fixing $\rho_c$,
the quantities $M,J,\Omega$ are functions of just one variable.
Using Eq.~\eqref{eq:dMdrho_J} we can test the stability condition
with only 3 configurations: those corresponding to $(\rho_c,\Omega)$,
$(\rho_c+\mathrm d\rho_c,\Omega)$, and $(\rho_c,\Omega+\mathrm d\Omega)$.
\footnote{One can use Eqs.~\eqref{eq:dMdrho_J}--\eqref{eq:dMdJ_rho}
also with the RNS code \citep{RNS, Stergioulas:1994ea}, using $\epsilon_c=\epsilon(\rho_c)$ (energy density at the center)
and $a\equiv r_{\rm surf}(\theta=0)/r_{\rm surf}(\theta=\pi/2)$
(stellar quasi-isotropic radii ratio) as independent variables
instead of $\rho_c$ and $\Omega$, respectively.}
Up to our knowledge this is the first time that Eq.~\eqref{eq:dMdrho_J}
is discussed and used; we check it in the third panel of Fig.~\ref{fig:search},
where the solid red line refers to the value of $\left.\partial M/\partial\rho_c\right|_J$ obtained
through Eq.~\eqref{eq:dMdrho_J} and the dashed light blue line is obtained 
with a root-finding cycle on $J$.
The two approaches agree within numerical uncertainties.

\subsection{Disk formation condition}
\label{ssec:disk}

We use the specific angular momentum $j$ of a fluid element 
(i.e., per baryon rest mass;  e.g. Eq.~(3.85) in
\citealp{Gourgoulhon:2011})
\begin{equation}
j=hW R^2 U^\phi.
\end{equation}
The co-rotating ISCO specific angular momentum for a Kerr black hole 
with total gravitational mass $M_{\rm BH}$ and angular momentum
$J_{\rm BH}$ is given by \citep[Eqs.~(2.12), (2.13), and (2.21) of][]{Bardeen+1972}
\begin{align}
\left.j\right._\mathrm{ISCO}={}&\sqrt{M_{\rm BH}}\frac{r_\mathrm{ISCO}^2-2\chi
M_{\rm BH}^{\frac32}r^{\frac12}_\mathrm{ISCO}+\chi^2 M_{\rm BH}^2}{d},
\end{align}
with 
\begin{align}
\chi={}&\frac{J_{\rm BH}}{M_{\rm BH}^2},\\
Z_1 ={}& \sqrt[3]{1-\chi ^2} \left(\sqrt[3]{1-\chi }+\sqrt[3]{\chi +1}\right)+1,\\
Z_2 ={}& \sqrt{3 \chi ^2+Z_1^2},\\
d={}&r_\mathrm{ISCO}^{\frac34}\sqrt{r^{\frac32}_\mathrm{ISCO}-3M_{\rm BH}r_\mathrm{ISCO}^{\frac12}+2\chi M_{\rm BH}^{\frac32}},\\
r_{\rm ISCO} ={}& M_{\rm BH} \left(3+Z_2-\sqrt{(3-Z_1) (Z_1+2 Z_2+3)}\right).
\end{align}

This allows to write the condition for disk formation as \citep{shapiro04,Margalit:2015qza}
\begin{equation}
\label{eq:disk}
j(r_\mathrm{surf},\pi/2) > \left.j\right._\mathrm{ISCO}(M_{\rm BH}=M,J_{\rm BH}=J),
\end{equation}
where $M$ and $J$ are the neutron star 
gravitational mass and total angular momentum, and 
$r_\mathrm{surf}$ is the stellar radius.
In Fig.~\ref{fig:theta_r_j} we plot the specific angular momentum distribution
in a neutron star. The black line corresponds to the ISCO angular momentum and
divides the material which will collapse
into the forming black hole and that will form a debris disk.
We remark that, to be fully consistent, we should have taken a black hole with a total
mass and angular momentum equal to that of the pre-collapse neutron star \emph{without}
the contribution from the debris disk. This could be accomplished within an
iterative procedure \citep{shapiro04}.
However, such a procedure is not well defined since the
local energy is not well defined in General Relativity and hence in
a neutron star. In any case we have checked that
the results obtained with the iterative procedure are indistinguishable
from those obtained without it because the debris disk has very
little mass and angular momentum (see discussion below).

\subsection{Parameter space exploration}
\label{ssec:search}

\begin{figure}
\includegraphics[width=\columnwidth]{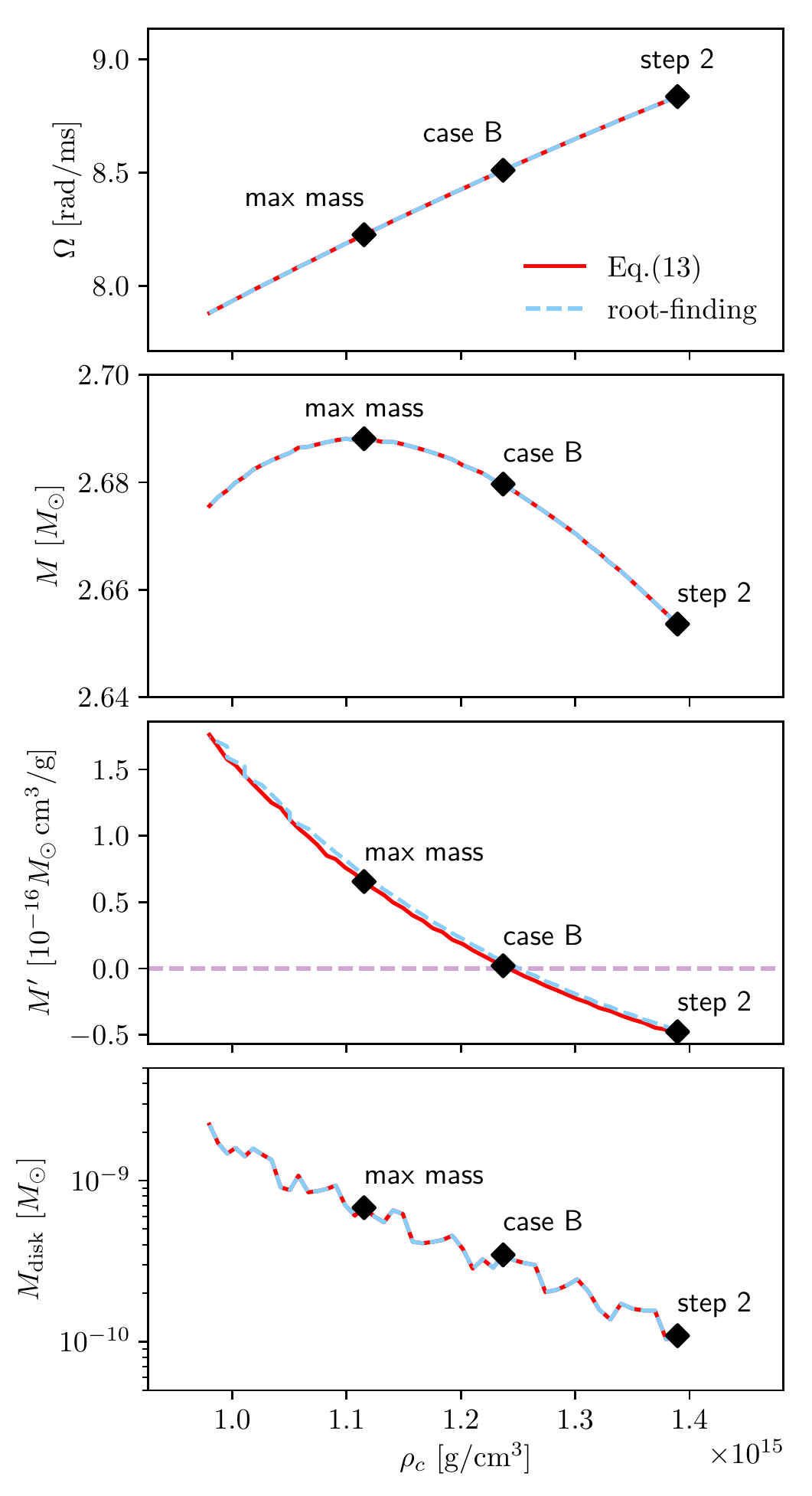}
\caption{Initial configuration search (extended to lower densities on the Keplerian line) for
the piecewise polytropic EOS corresponding to Case B (see Tab.~\ref{tab:cases}).
The configurations discussed in the text are marked.
From top to bottom we show the angular frequency $\Omega$, the
gravitational mass $M$,
the partial derivative at constant $J$: $\left.M'\equiv\partial M/\partial\rho_c\right|_J$,
and the baryonic mass of the disk
estimated from the equilibrium configuration $M_{\rm disk}$
[i.e., the baryon mass of the material that fulfill Eq.~\eqref{eq:disk}], all plotted against the central density $\rho_c$.
The solid red lines are obtained using Eq.~\eqref{eq:dMdrho_J}
and the dashed light blue lines are obtained with a root-finding cycle on $J$.
The results of the two approaches differ only in $M'$, in which case are consistent within
the numerical uncertainties.}
\label{fig:search}
\end{figure}

\begin{figure*}
\includegraphics[width=1.5\columnwidth]{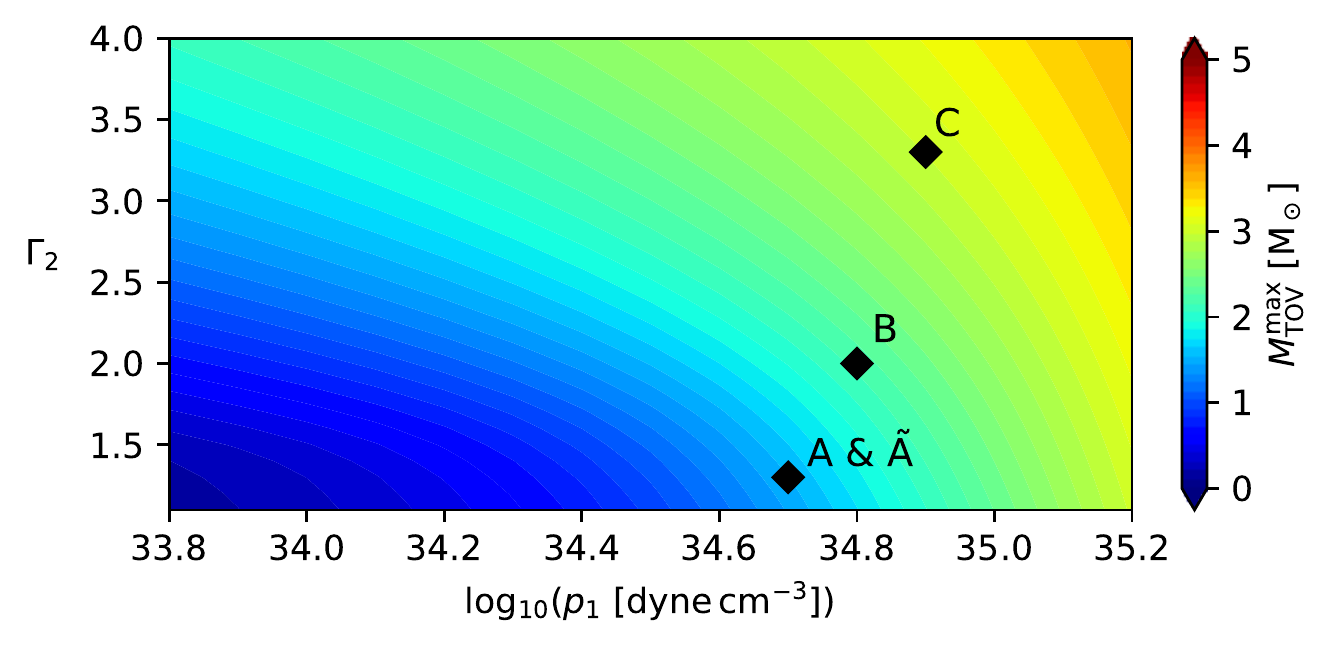}
\includegraphics[width=1.5\columnwidth]{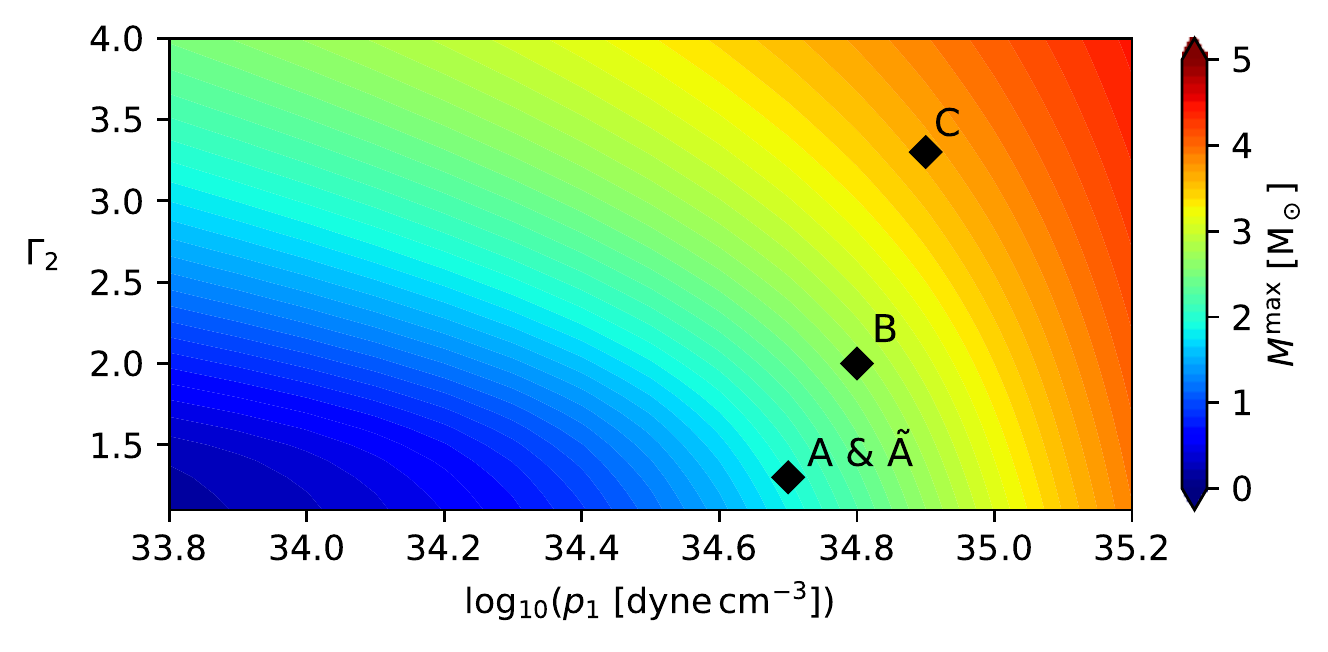}
\includegraphics[width=1.5\columnwidth]{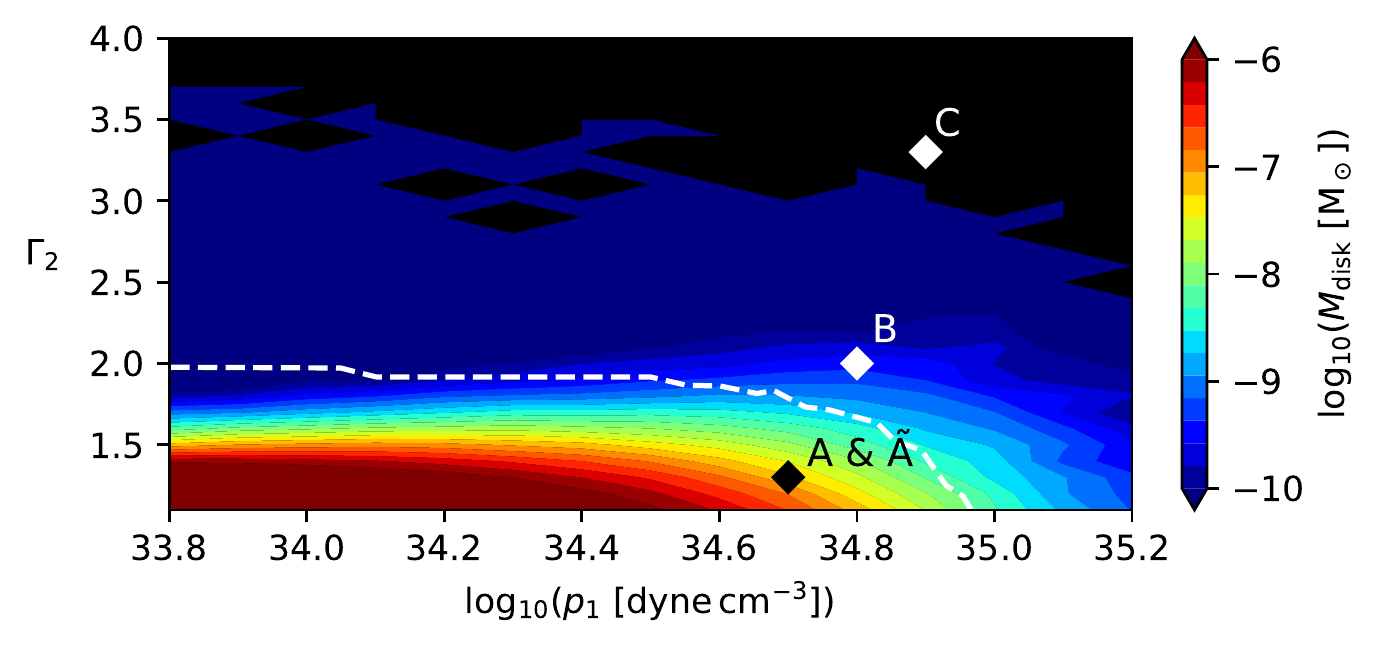}
\caption{Stellar properties dependence on the EOS parameters $p_1$ and $\Gamma_2$.
From top to bottom:
maximal TOV gravitational mass, maximal gravitational mass (at the Keplerian frequency),
and estimated disk baryon mass of the marginally stable Keplerian configuration.
The black area in the lower plot represents the parameter space for which no disk
formation is possible ($M_{\rm disk}=0$). We mark with diamonds the configurations
that we have further studied with dynamical simulations.
The white dashed line denotes the disk/no-disk separation curve previously
found by \citet{Margalit:2015qza}.
}
\label{fig:disk_mass}
\end{figure*}

Following the above discussion we estimate, for each choice of
the EOS (specified by parameters $p_1$ and $\Gamma_2$), the baryonic mass of the disk that can be generated
from the collapse of the maximally rotating (Keplerian) configuration.
We remark that in general the marginally stable Keplerian configuration,
which is the stable configuration with greater angular velocity,
\emph{is not} the configuration with maximal gravitational mass
at the Kepler frequency (see e.g. Fig.~2 in \citealp{Stergioulas:1994ea}
or Fig.~\ref{fig:search} of this paper).

The code settings adopted in the parameter space exploration are the same
of the ``baseline'' configuration described in Appendix~\ref{app:XNS_vs_RNS},
apart for the radius of the inner grid which is set to $20$ (in
code units, see Appendix~\ref{app:XNS_vs_RNS} for details).
Our procedure to find the interesting stellar models,
for each choice of the EOS parameters $p_1$ and $\Gamma_2$, is:
\begin{enumerate}[(1)]
\item start from the maximal TOV (spherical) mass configuration, with $\Omega=0$ and central density
$\rho_c$.
\item keep $\rho_c$ fixed and increase $\Omega$ until the Keplerian (i.e., maximally rotating) 
configuration is reached.
We will refer to this configuration as ``step 2''\footnote{For 
most of the EOS parameter space
considered in this paper, the step 2 configuration is unstable ($\left.\partial M/\partial \rho_c\right|_J<0$)
and has a central density greater than that of
the maximal mass configuration (see Fig.~\ref{fig:search}). When this does not hold,
for very low $p_1$ and $\Gamma_2$, we have re-started the search from a greater central density.}.
The Keplerian angular speed $\Omega_\mathrm{K}$ is determined by evaluating
the co-rotating case of Eq.~(4.93) of 
\citet{Gourgoulhon:2011} at the equator,
\begin{equation}
\label{eq:keplerian}
\Omega_\mathrm{K}=\omega + \frac{\omega'R}{2R'}+\sqrt{\frac{N'N}{R'R}+\left(
\frac{\omega'R}{2R'}\right)^2},
\end{equation}
where all quantities are evaluated at the equator and primes denote derivatives
along the radial direction $r$.
\item repeat:
\begin{enumerate}[(a)]
   \item compute $\left.\partial M/\partial \rho_c\right|_J$
   [Eq.~\eqref{eq:dMdrho_J}] varying $\rho_c$ and $\Omega$.
   \item if
   $\left.\partial M/\partial \rho_c\right|_J \geq 0$, the configuration is 
   maximally rotating and at the verge of collapse (i.e., it is the marginally stable Keplerian configuration).
   \item if the mass is lower than that of the previous configuration,
   the previous is the maximal mass one.
   \item once the maximal mass configuration and the marginally stable Keplerian configuration
   are found, exit the cycle.
   \item reduce $\Omega$ by $\rm d\Omega = \unit[10^{-4}]{M_\odot^{-1}}$ ($c=G=1$ units).
   \item reduce $\rho_c$ until you reach the Keplerian configuration.
\end{enumerate}
\end{enumerate}
In general, the step~2 and the maximal mass configurations can be stable or
unstable \citep{Stergioulas:1994ea}.
As an example, in Fig.~\ref{fig:search} we report the search for a case in which the
maximal mass configuration is stable and the step~2 configuration
is unstable, which is the most common case in our analysis.

We estimate the mass of the debris disk from a given equilibrium configuration
integrating the baryon mass in the neutron star that fulfills the condition
$j(r,\theta)>\left.j\right._\mathrm{ISCO}$ (lower plot of Fig.~\ref{fig:disk_mass}).
This estimated disk mass increases as we step to lower central densities along
the Keplerian curve (see lower plot in Fig.~\ref{fig:search}). However, these
lower density configurations would not give rise to a debris disk because they
are stable.

We have explored the same EOS parameter space as
\citet{Margalit:2015qza}, namely $p_1\in [ 10^{33.8}; 10^{35.2}]\ 
\unit{dyne\,cm^{-3}}$ and $\Gamma_2\in[1.1; 4]$, our results
are shown in Fig.~\ref{fig:disk_mass}.  Most of the EOS parameter
choices result in disk formation;  moreover the general trend is that 
the greater the maximal mass (stiffer EOSs, i.e., greater $p_1$ and 
$\Gamma_2$), the smaller the disk mass of the marginally stable Keplerian configuration.

We select a few configurations that we also study
with fully dynamic Numerical Relativity simulations (Cases
A, \~A, B, and C, see marks in Fig.~\ref{fig:disk_mass}). These
configurations are reported in Tab.~\ref{tab:cases}. The rationale behind
our choices is the following:
\begin{itemize}
\item[\textbf{A}] marginally stable Keplerian configuration with the
greatest disk mass compatible with the request
to have a maximal TOV mass greater than $1.5{M_\odot}$,
see upper plot in Fig.~\ref{fig:disk_mass}.
The corresponding value of $\Gamma_2= 1.3$ is substantially smaller
than what is considered realistic ($\ge 2.5$; see e.g. Fig.~5 in \citealp{rosswog02a} for
an illustration).
\item[\textbf{\~A}] maximal mass configuration equivalent to another
Case studied with a dynamical simulation that results in a collapse to test
our prediction on the stability of
the maximal mass configuration (we picked a stable case).
\item[\textbf{B}] marginally stable Keplerian configuration with
the maximal TOV mass equal to $2{M_\odot}$
and $\Gamma_2=2$. This choice of the high density polytropic exponent
is still smaller than current predictions but it is a common choice
in many numerical applications.
\item[\textbf{C}] marginally stable Keplerian configuration with
no predicted disk and a maximal TOV mass smaller than $\unit[3]{M_\odot}$.
\end{itemize}

\begin{table*}
  \centering    
  \caption{Properties of the configurations marked in Fig.~\ref{fig:disk_mass}
  and described in Sec.~\ref{ssec:search}. The columns contain, in order:
  the configuration name (``Case''); the EOS parameters ($p_1$ in $\rm dyne\,cm^{-3}$ and $\Gamma_2$);
  the configuration parameters (central density $\rho_c$, angular speed
  seen by an observer at infinity $\Omega$, axes ratio $a$, gravitational mass $M$); 
  and the disk baryon mass
  (both determined from the equilibrium configuration $M_{\rm disk}^{\rm eq}$ and
  from the dynamical simulation about $8\,{\rm ms}$
  after its beginning $M_{\rm disk}^{\rm dyn}$). Note that Case \~A
  does not collapse, as expected.
}
  \begin{tabular}{c|cc|cccc|ccc}        
    \hline
    Case    & $\log_{10}(p_1)$ & $\Gamma_2$ & $\rho_c\ [\rm 10^{15}\,g\,cm^{-3}]$ & $\Omega\ [\rm rad/ms]$ & $a$ &
    $M\ [\rm M_\odot]$ & $M_{\rm disk}^{\rm eq}\ [\rm M_\odot]$ & $M_{\rm disk}^{\rm dyn}\ [\rm M_\odot]$ \\ 
    \hline
    A   & 34.7 & 1.3 & 1.212 & 7.016 & 0.5567 & 1.924 & $7\times10^{-8}$ & $\lesssim 10^{-7}$ \\
    \~A & 34.7 & 1.3 & 1.033 & 6.834 & 0.5524 & 1.933 & - &  - \\
    B   & 34.8 & 2.0 & 1.238 & 8.511 & 0.5599 & 2.679 & $3\times10^{-10}$ &  $\lesssim 10^{-7}$ \\
    C   & 34.9 & 3.3 & 0.9634 & 9.573 & 0.5630 & 3.614 & $4\times 10^{-13}$ & $\lesssim 10^{-8}$ \\    
    \hline
  \end{tabular} 
 \label{tab:cases}
\end{table*}

\subsection{Comparison with Margalit et al.~(2015)}
\label{ssec:Margalit}
While in good qualitative agreement,  
our results differ quantitatively from those of \citet{Margalit:2015qza}.
In fact, we find that disk formation is possible also for $\Gamma_2\gtrsim 2$ and 
$\log_{10}(p_1/\unit{dyne\,cm^{-2}})\gtrsim 34.9$, cf.~lower panel of Fig.~\ref{fig:disk_mass}.
None of the cases considered in our study, however, is a candidate for producing 
an sGRB, because the disk mass is too small, cf.~Eq.~\eqref{eq:mdisk}.

In principle, the main difference in the employed methods 
between our work and \citet{Margalit:2015qza} are:
\begin{itemize}
\item \citet{Margalit:2015qza} uses the RNS code, while we use XNS.
As we show in Appendix~\ref{app:XNS_vs_RNS}, the configurations found by both codes
are in very good agreement.
\item \citet{Margalit:2015qza} searches for the maximal mass configuration 
instead of the marginally stable configuration.
As argued in Sec.~\ref{ssec:search} and in \citet{Stergioulas:1994ea},
the maximal mass configuration is not necessarily unstable,
because the stability condition should be checked at constant $J$,
and in any case it is not on the verge of instability.
However, we checked that the maximal mass configuration
generates a disk similar to that of the marginally stable Keplerian
configuration.
\end{itemize}
We note that the step~2 configuration actually
reproduces Fig.~2 of \citet{Margalit:2015qza}. However, this configuration is
unstable
for most of the choices of the EOS parameters (apart for very small $p_1$ and $\Gamma_2$)
and was therefore not considered in our analysis.

\subsection{Fit to realistic EOSs of Read et al. (2009)}
\label{ssec:ReadEOSs}

In addition to the general consideration of neutron stars described by a 
2-piece polytropic core, we have applied the outlined procedure to some more 
realistic multi-piecewise polytropic EOSs.  
Those fits have been constructed in \citet{Read:2008iy} and model 
EOSs describing full tabulated EOSs for different nuclear physical models. 
The results for these EOSs are given in Tab.~\ref{tab:ReadEOSs}.
We have chosen this subset of EOSs since it is in agreement with current observations: 
(i) maximum supported masses are above $2.0M_\odot$ \citep{Antoniadis:2013pzd};
(ii) maximum supported masses are below  $\sim 2.3M_\odot$ \citep{Rezzolla:2017aly,Shibata:2017xdx,
Ruiz:2017due,Margalit:2017dij}; 
and (ii) the compactness and tidal deformability 
are in agreement with the measurements obtained from GW170817 \citep{abbott17b,Abbott:2018wiz}.
The results for realistic EOSs confirm the conclusions for the EOS
parameter search made in this section, namely that even if a debris disk
can form, its mass is too small to generate an energetic GRB.\\

An important point to stress here is that we are discussing the mass and the
extractable GRB energy of a debris disk formed by material of the pre-collapse
neutron star. This means that we are not addressing the possibility that
the GRB is caused by a pre-existing debris disk \citep{Michel+Dessler1981}, for example due to fallback
from the original supernova event. These disks may potentially be more massive than
the disks we predict in our analysis (e.g., \citealp{Wang+2006} found observational
evidence of a fallback disk of $\sim\unit[10^{-5}]{M_\odot}$, see also \citealp{Wang2014} for a recent review).

\begin{table}
  \centering    
  \caption{Disk formation for some realistic EOSs (see text for details), whose piecewise polytropic
  fit is given in \protect\citet{Read:2008iy} (with an SLy crust at low densities).  We report the EOS name (1st
  column), the gravitational mass of the maximal mass configuration (2nd
  column), the angular velocity and disk baryonic
  mass estimated from the equilibrium model of the marginally stable Keplerian configuration (3rd and 4th columns
  respectively).
  We remark that the angular velocity of the marginally stable Keplerian configuration
  is also the maximal one for stable configurations, see
  Fig.~\ref{fig:search}.
}

  \begin{tabular}{cccc}        
    \hline
    EOS    & $M^{\rm max}\ [\rm M_\odot]$ & $\Omega\ [\rm rad/ms]$ & $M^{\rm eq}_{\rm disk}\ [\rm M_\odot]$ \\ 
    \hline
    SLy  & 2.415 & 11.50 & $5\times10^{-13}$ \\
    APR4 & 2.594 & 12.32 & $2\times10^{-12}$ \\
    WFF1 & 2.534 & 13.54 & $8\times10^{-13}$ \\
    WFF2 & 2.604 & 12.67 & $1\times10^{-13}$ \\
    ENG  & 2.656 & 11.52 & $3\times10^{-12}$ \\
    ALF2 & 2.399 & 9.187 & $2\times10^{-10}$ \\
    \hline
  \end{tabular} 
 \label{tab:ReadEOSs}
\end{table}

\section{Dynamical Evolutions}
\label{sec:evolution}

In the following we study the configurations marked in
Fig.~\ref{fig:disk_mass} and described in Tab.~\ref{tab:cases} and
Sec.~\ref{ssec:search} to determine whether dynamical effects can facilitate
the debris disk formation.

\subsection{Cases A and $\tilde{\rm A}$}

\begin{figure}
\includegraphics[width=\columnwidth]{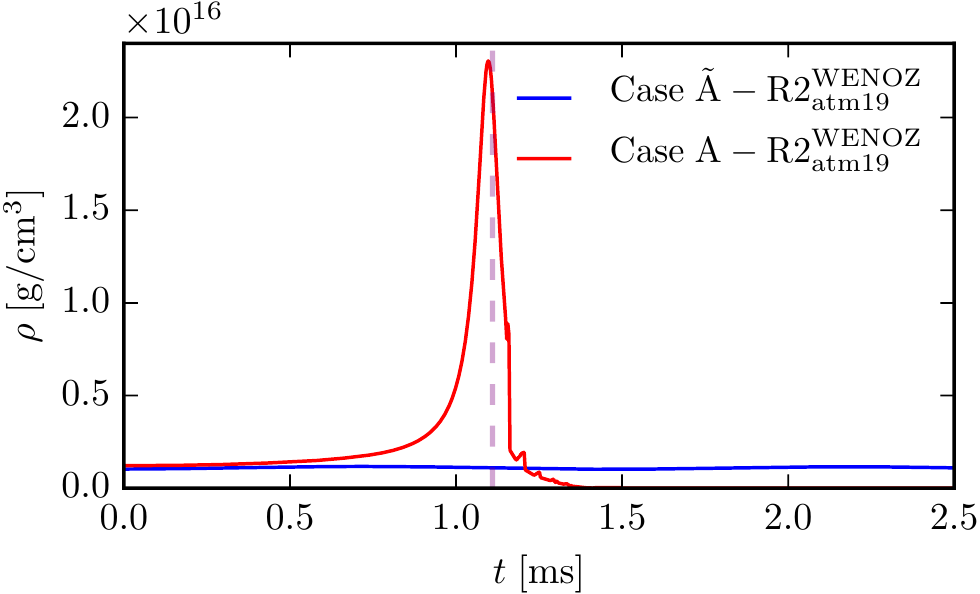}
\caption{Maximum density for the maximal mass configuration
(Case~\~A, blue line) and marginally stable Keplerian configuration
(Case A, red line). We mark the formation of the 
apparent horizon for Case A with a vertical purple dashed line.}
\label{fig:CaseA_MRMM_MRAU}
\end{figure}

Cases A and \~A employ an EOS with $\log(p_1)=34.7$ and $\Gamma_2=1.3$, 
see Tab.~\ref{tab:cases}. The Case A (marginally stable Keplerian
configuration) is characterized by a central density of $\rho_c=1.212\times
10^{15}\ {\rm g/cm^3}$ 
and an angular speed of $\Omega=\unit[7.016]{rad/ms}$. 
The Case \~A (maximal mass configuration) has a central density $\sim 15\%$ lower than
Case A and an angular speed of $\Omega=\unit[6.834]{rad/ms}$. 
Within our simulations we trigger the gravitational collapse by introducing 
a small artificial pressure perturbation. This is a common approach for the 
study of gravitational collapse. Generally, large perturbations lead to a faster 
collapse which reduces the computational cost of the individual simulations, but 
on the other hand it might affect the dynamical evolution. 
We employ a small perturbation of $0.05\%$ to reduce nonphysical 
effects\footnote{We also performed a subset of simulations 
with different pressure perturbations and find consistent results.},
reminding the reader that the introduced pressure 
perturbation leads to Hamiltonian constraint violations at $t=0$. 
Previous studies have used larger perturbations, 
see e.g.~\citet{Giacomazzo:2012bw} where a $0.1\%$ perturbation, 
\citet{Dietrich:2014wja} where a $0.5\%$ perturbation, and 
\citet{Baiotti:2004wn,Baiotti:2007np,Reisswig:2012nc} where 
a $2\%$ perturbation were applied. 

\paragraph*{Comparison of maximal mass and marginally stable Keplerian configurations.}
Before discussing the gravitational collapse in detail, 
we compare the simulations of the maximal mass and 
the marginally stable Keplerian configurations. 
Figure~\ref{fig:CaseA_MRMM_MRAU} shows the maximum density 
evolution for both cases, where the maximal mass configuration is shown in blue
(Case \~A)
and the marginally stable Keplerian configuration in red (Case A). 
We find that, as outlined in our previous discussion 
(Sec.~\ref{ssec:stability}), the maximal mass configuration is stable 
and does not undergo gravitational collapse while the 
marginally stable Keplerian configuration is characterized by a rapid increase of the 
central density until a BH forms at $\sim 1\ {\rm ms}$
after the begin of the simulation. The vertical dashed line  in Fig.~\ref{fig:CaseA_MRMM_MRAU}
denotes the horizon formation.
For the maximal mass model we find small density oscillations
introduced by the pressure perturbation (not visible at the density scale
of Fig.~\ref{fig:CaseA_MRMM_MRAU}). Whereas these oscillations are too 
small to cause a gravitational collapse, imposing a larger pressure
perturbation would have led 
to BH formation also for the maximal mass configuration. 

\begin{figure}
\includegraphics[width=0.49\textwidth]{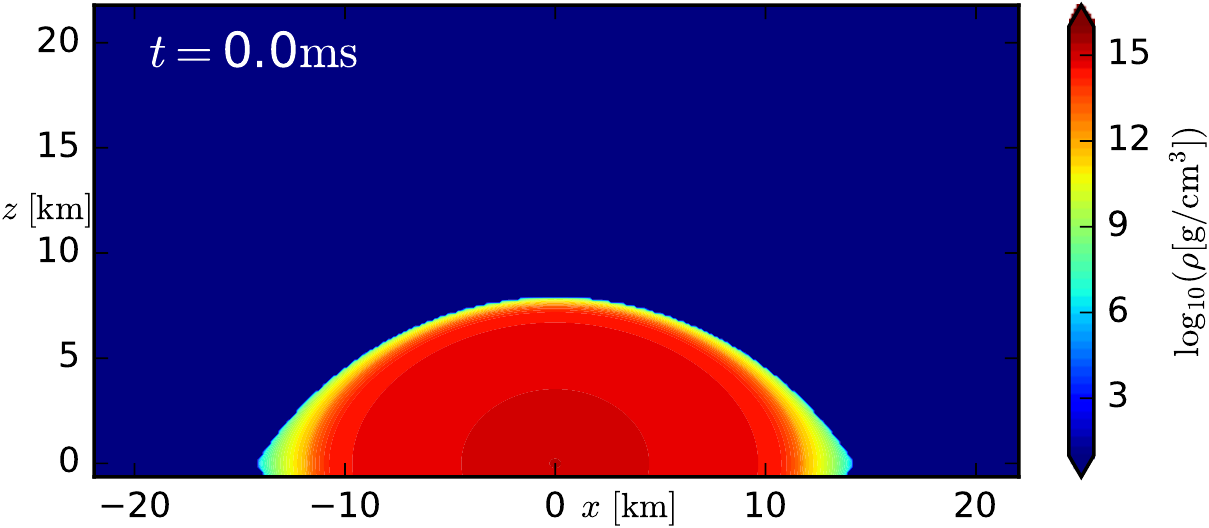} 
\includegraphics[width=0.49\textwidth]{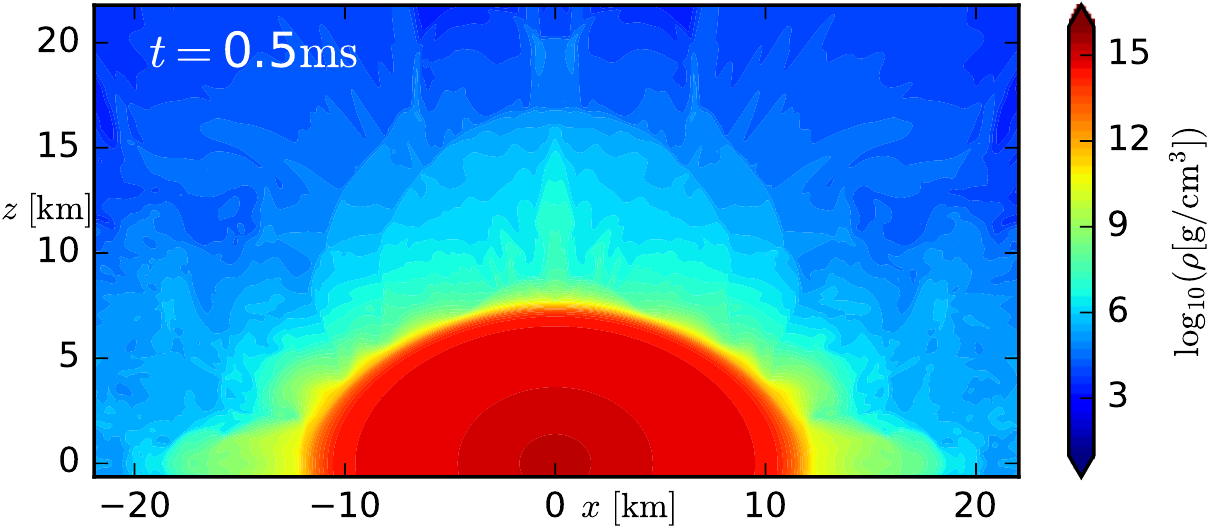} 
\includegraphics[width=0.49\textwidth]{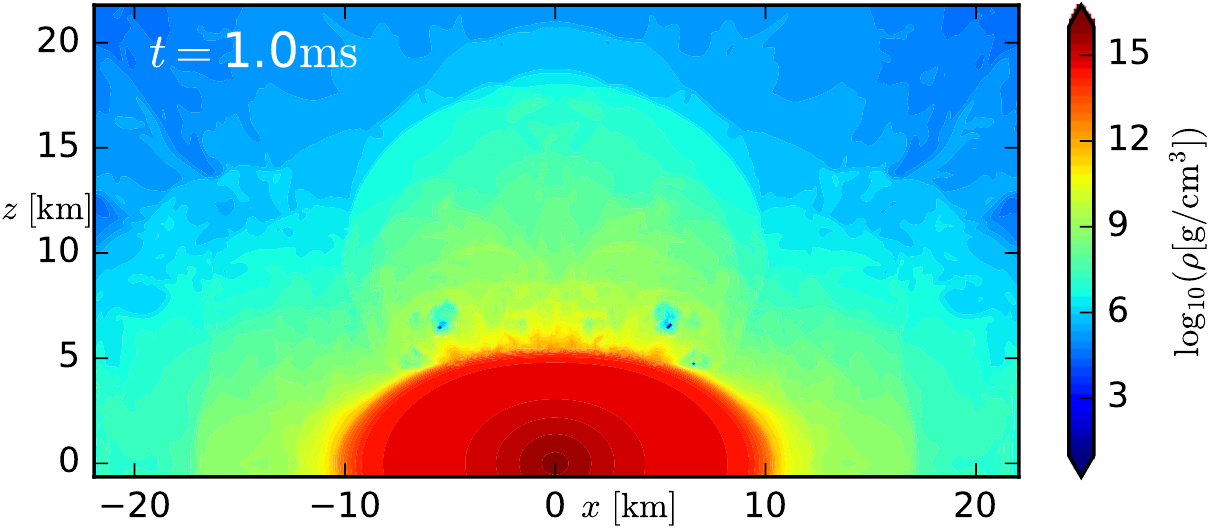} 
\includegraphics[width=0.49\textwidth]{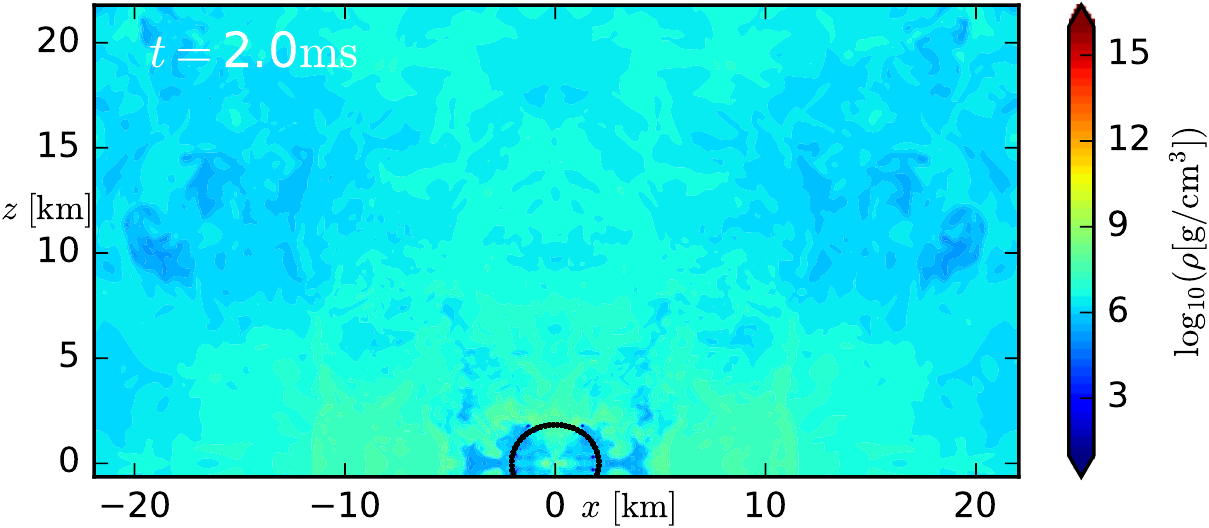} 
\includegraphics[width=0.49\textwidth]{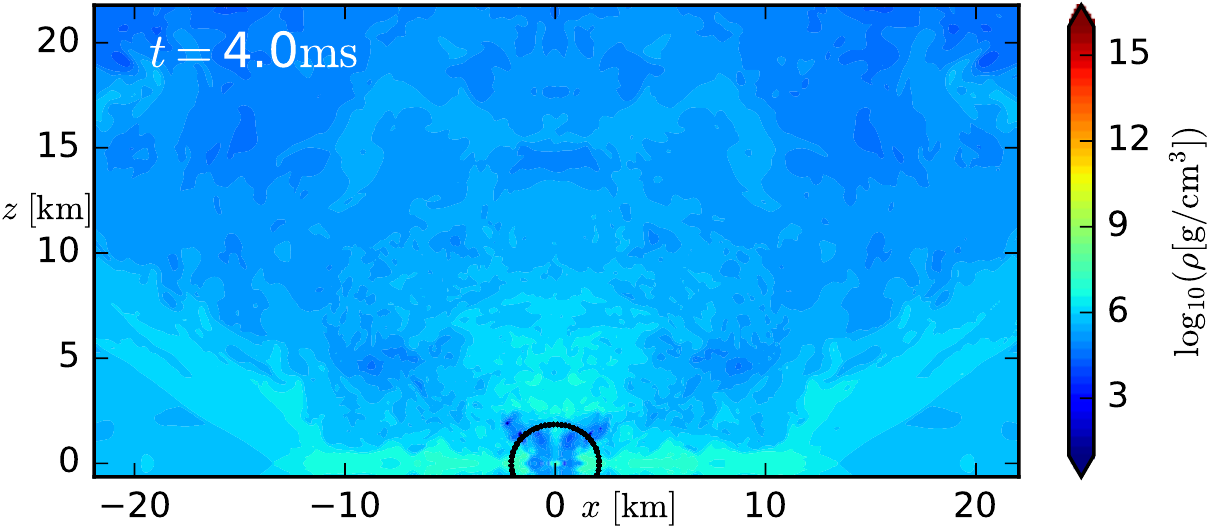} 
\includegraphics[width=0.49\textwidth]{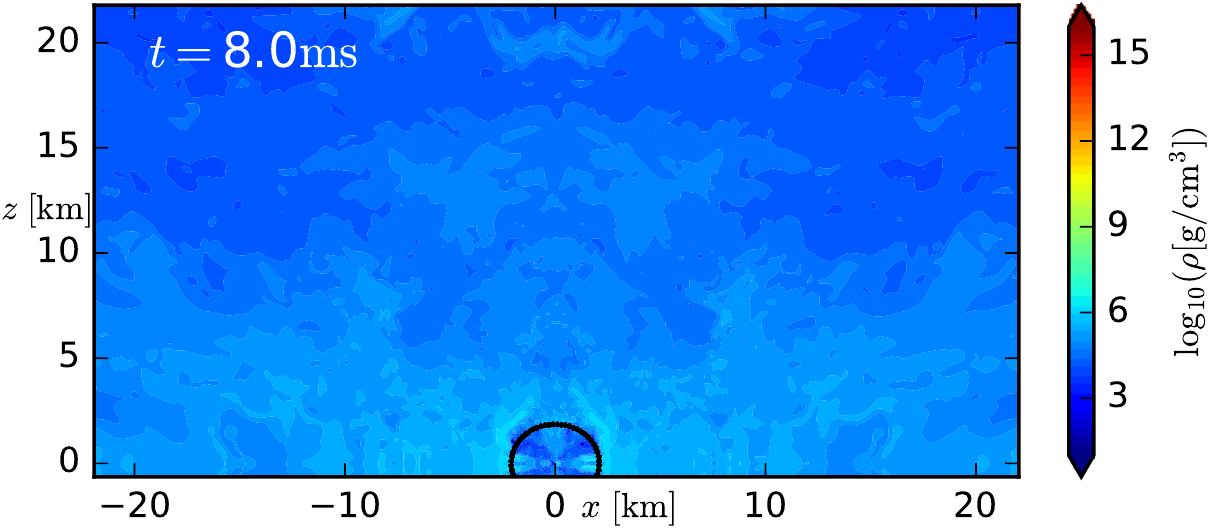} 
\caption{ 
Snapshots of the density profile in the x-z-plane 
at $t=0,0.5,1.0,2.0,4.0,8.0\ {\rm ms}$ for the Case A.  
The snapshot times are marked 
in Fig.~\ref{fig:caseA} with black diamonds. The black solid line marks 
the BH horizon.}
\label{fig:evolution}
\end{figure}

\begin{figure}
\includegraphics[width=\columnwidth]{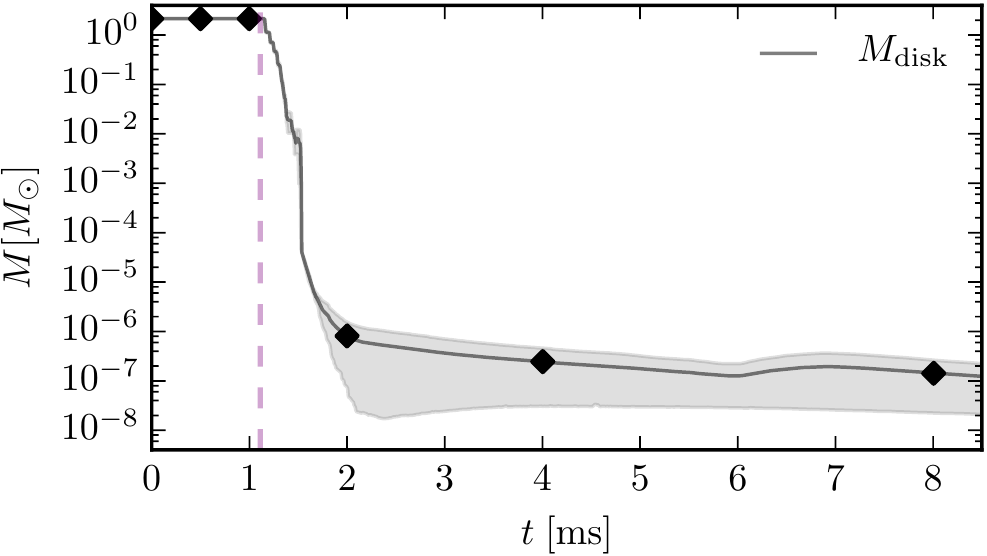}
\caption{
Evolution of the bound (disk) mass for Case A. 
Black diamonds refer to the times shown in Fig.~\ref{fig:evolution} and
the vertical dashed purple line to the formation time of the apparent horizon. 
The shaded region shows the difference between the two higher resolution simulations.}
\label{fig:caseA}
\end{figure}

\paragraph*{Collapse morphology.}
In the following we discuss the dynamics during the gravitational collapse 
of the marginally stable Keplerian model. For this purpose we show for different 
instants of times the density within the x-z-plane 
(corresponding to a slice with constant $\phi$ in the
equilibrium case) in Fig.~\ref{fig:evolution}. 
The shown time snapshots are marked in Fig.~\ref{fig:caseA},
where we report the disk mass estimated as average of the two highest resolutions 
(the shaded region shows the difference between these simulations). 
The top panel of Fig.~\ref{fig:evolution} shows the initial equilibrium configuration 
at $t=0\ {\rm ms}$. 
The stellar shape is characterized by its oblate form due to the large intrinsic 
rotation. 

At $t=0.5\ {\rm ms}$ (second panel of Fig.~\ref{fig:evolution}) the stellar surface is 
less sharp compared to the initial configuration. This is typically observed 
in all Numerical Relativity simulations 
of neutron star spacetimes using grid-based codes, 
see e.g.~\citet{Guercilena:2016fdl} for further discussions. 
It is introduced by the fact that the numerical scheme is unable to resolve the sharp, 
step-like surface of the star. 
This effect becomes even more pronounced due to artificial 
shock heating at the stellar surface.
While an increased resolution and less dissipative 
schemes for the numerical fluxes reduce the effect, there are currently 
no full 3D Numerical Relativity simulations 
of dynamical spacetimes which retain the exact shape of the surface. 
We suggest that due to this effect Numerical Relativity simulations 
are likely to overestimate the material surrounding the NS.

At $t=1.0\ {\rm ms}$ (third panel of Fig.~\ref{fig:evolution}) the 
star has further contracted and the central density 
has increased. Most notably some low density material 
leaves the star with high velocity along the z-axis. 
This matter becomes unbound and is ejected from the system. 
We mark material as unbound/ejected once the geodesic criterion is fulfilled, 
i.e., when the time-component of the four-velocity is $u_t<-1$ 
and when the radial component of the velocity is positive. 
The ejection of material is caused by shocks at the stellar surface. 
Since those shocks might be associated to the artificial atmosphere 
employed in the dynamical evolution, we assess the error of the numerical method
by simulating configurations with different resolutions and atmosphere values, 
as well as flux limiting schemes, see Fig.~\ref{fig:caseAres} and the discussion below. 

At $t=2\ {\rm ms}$ (forth panel of Fig.~\ref{fig:evolution}), 
the star has collapsed to a BH; we mark the apparent horizon with a 
black solid line. The density dropped several orders of magnitude 
and reaches now maximum values of $\rho \sim 10^7 {\rm g/cm^3}$. 
At this time the bound mass (namely, the debris disk mass) has decreased
to $M_{\rm disk} \approx  \unit[10^{-6}]{M_\odot}$. 

At $t=4\ {\rm ms}$ (fifth panel of Fig.~\ref{fig:evolution}), 
the density decreases further to $\rho \sim 10^6 {\rm g/cm^3}$
and finally at $t=8\ {\rm ms}$ (last panel) the density surrounding 
the central BH has dropped to $\rho \lesssim 10^5 {\rm g/cm^3}$.
The final disk mass at this time has settled at about 
$\sim \unit[10^{-7}]{M_\odot}$.

\begin{figure*}
\includegraphics[width=0.98\columnwidth]{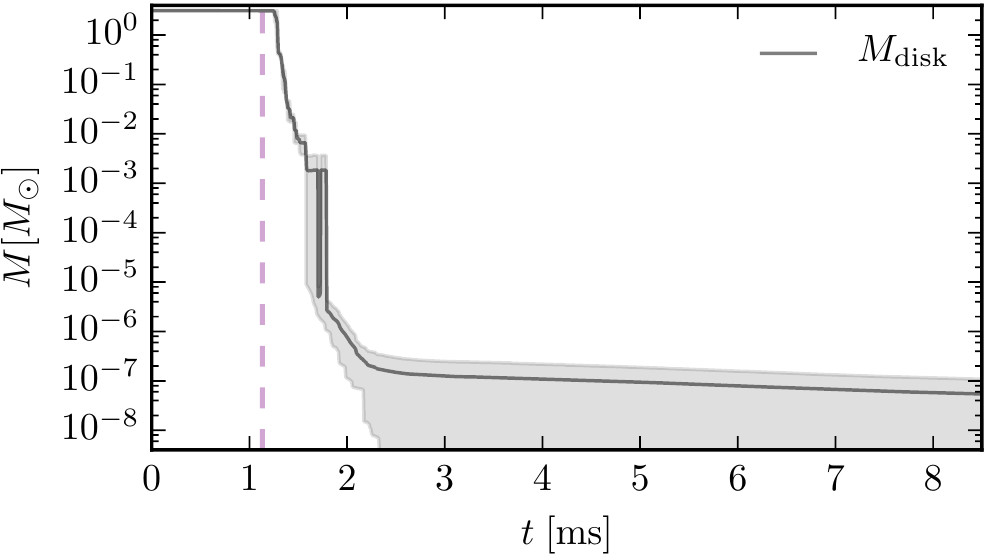}
\includegraphics[width=0.98\columnwidth]{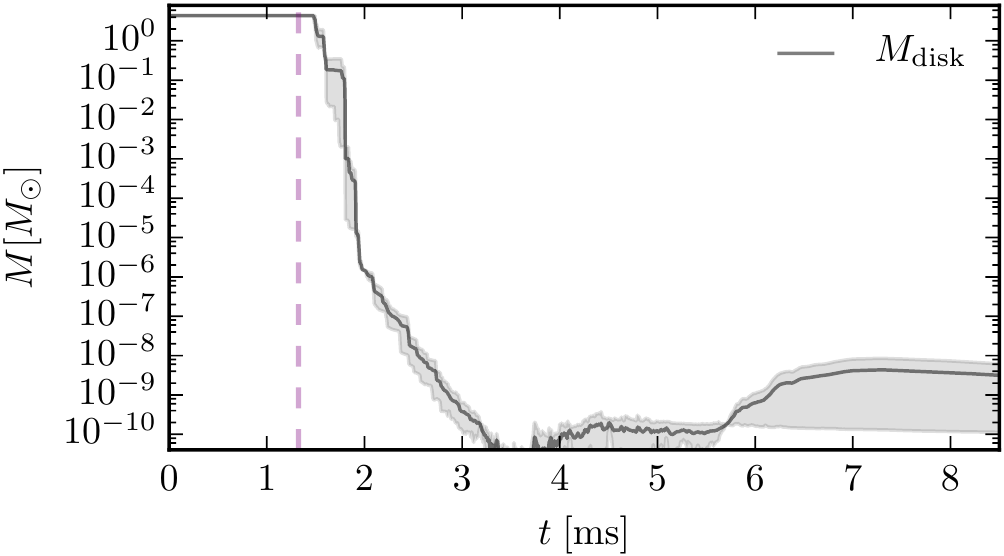}
\caption{
Evolution of the bound (disk) mass for Case B (left panel) and Case C (right panel). 
The vertical dashed, purple lines mark the formation time of an apparent horizon and 
the shaded region represents the differences between the two higher resolution simulations.}
\label{fig:caseBC}
\end{figure*}

\subsection{Case B}

For Case B we employ an EOS characterized by 
$\log(p_1)=34.8$ and $\Gamma_2=2.0$. 
The marginally stable Keplerian configuration has a central 
density of $\rho_c = 1.238\times10^{15}\ {\rm g/cm^3}$ 
and an angular velocity of $\Omega = \unit[8.511]{rad/ms}$. 
Since the collapse dynamics follows the same qualitative steps 
outlines for Case A in Fig.~\ref{fig:evolution}, 
we restrict our considerations to quantitative statements.
The debris disk mass is slightly smaller than for Case A, 
as expected from our findings for the equilibrium configuration,
but overall also of the order of $\sim \unit[10^{-7}]{M_\odot}$. 
Similarly to Case A we obtain an ejecta mass 
of the order of $\unit[10^{-4}]{M_\odot}$; 
see discussion below for more details. 

\subsection{Case C}

Case C employs an EOS determined by 
$\log(p_1)=34.9$ and $\Gamma_2=3.3$. 
The marginally stable Keplerian configuration has a central 
density of $\rho_c =9.634\times10^{14}\ {\rm g/cm^3}$ 
and an angular speed of $\Omega = \unit[9.573]{rad/ms}$. 
We find for this setup that the mass of the disk (bound material) falls below 
$10^{-10}\ M_\odot$ about $2\ {\rm ms}$ after BH formation. 
The increase of bound mass after this time might be caused
either by material which is initially marked as unbound and later falls back onto 
the remnant\footnote{It is possible that material which is first marked 
as unbound falls back onto the remnant, since the geodesic criterion used to 
characterize fluid elements assumes that fluid elements follow 
a geodesic motion, which is only approximately correct for a dynamical spacetimes 
as the one considered in this article.} or simply by inaccuracies of the numerical scheme. 
However, overall Case C produces the smallest amount of 
bound material as expected from the equilibrium configuration analysis.

\subsection{Accessing the numerical uncertainty}
\begin{figure}
\includegraphics[width=\columnwidth]{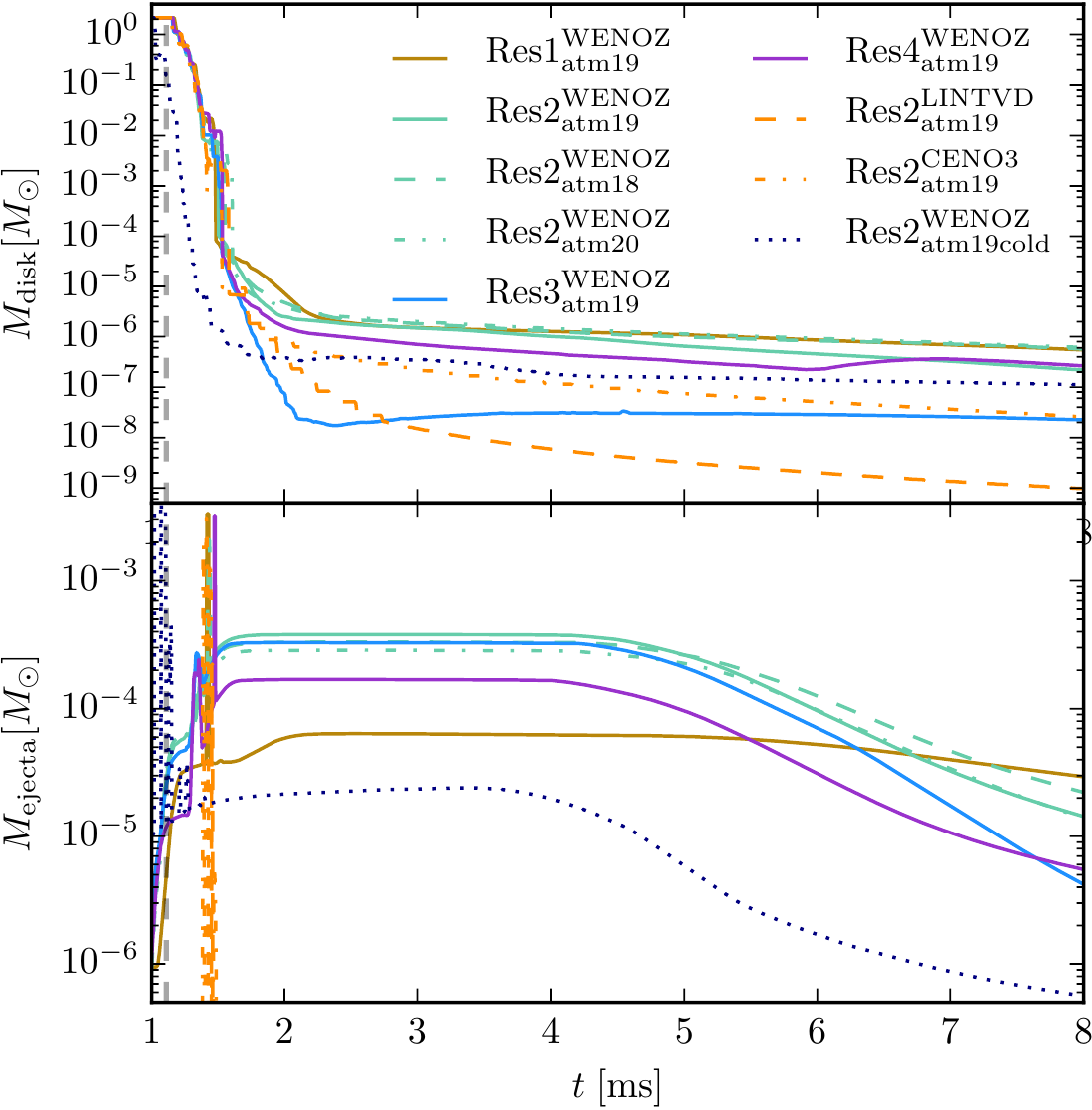}
\caption{Evolution of the bound (top panel) and the unbound 
(lower panel) baryonic mass for different resolutions, atmosphere settings, 
and numerical flux limiters. 
The vertical dashed line refers to the apparent horizon formation for setup 
${\rm Res4}_{\rm atm 19}^{\rm WENOZ}$. 
Note that the main reason for the decrease in the ejecta mass after about 
$\sim 5\ {\rm ms}$ (see Fig.~\ref{fig:caseA}) is the material which leaves the numerical 
domain covered by our simulation. }
\label{fig:caseAres}
\end{figure}

Fig.~\ref{fig:caseAres} gives an overview of the bound (disk) mass 
and the unbound (ejecta) mass for all simulations of Case A. 

\paragraph*{Artificial atmosphere.}
Let us start by considering the imprint of the artificial atmosphere, cf.~green lines
in top and bottom panels. Although the artificial atmosphere threshold has been varied
by a factor of $100$, we find that the disk and ejecta masses are almost unchanged. 
Therefore, although the artificial atmosphere 
introduces errors, the previous conclusions remain valid. 

\paragraph*{Resolution.}
We continue the discussion by focusing on the simulations with different resolutions. 
We have varied the resolution by a factor of three, which generally is a very large range for 
full Numerical Relativity simulations where computational costs scale with the forth power of the 
number of grid points. We do find that the results are not monotonically converging with increasing resolution.
This behavior is unfortunately often seen in full 
3D Numerical Relativity simulations estimating disk and ejecta masses, 
see e.g.~\cite{Hotokezaka:2012ze,Dietrich:2016fpt,Fujibayashi:2017puw}. 
However, although precise statement about the bound/unbound mass can not be made, 
the fact that the 
mass estimates change only about one order of magnitude
for the large range of resolutions employed leads to 
the conclusion that the order of magnitude estimates necessary 
for our study are indeed valid. 

\paragraph*{Numerical flux limiter.}
We also discuss the imprint of the flux limiter used in the GRHD scheme. 
For this purpose we employ 3 different flux reconstruction schemes:
LINTVD~\citep{Shu:1989}, CENO3~\citep{Liu:1998,DelZanna:2002rv}, 
and WENOZ~\citep{Borges:2008a,Bernuzzi:2012ci}.
As expected we find that less sophisticated, lower order schemes as LINTVD 
and CENO3 predict smaller bound and unbound masses. 
In particular the ejecta mass drops to zero for these two schemes. 
This analysis shows that high order flux limiters 
as WENOZ seem to be required for a proper modeling of the system.
Although we can not exclude that with even more 
improved HRSC methods larger disk masses might be observed, 
we do expect that the results are robust and allow order 
of magnitude estimates. This statement is based on 
investigations of binary systems that 
show that the WENOZ reconstruction scheme is 
among the state-of-the-art methods and allows accurate 
and reliable simulations of neutron star spacetimes, 
see e.g.~\citet{Bernuzzi:2012ci,Bernuzzi:2016pie}. 

\paragraph*{Thermal effects.}
Finally, we consider the imprint of the thermal effects added through Eq.~\eqref{eq:p_evolution}. 
For this purpose we compare the setups ${\rm Res2_{atm19}^{WENOZ}}$ and ${\rm Res2_{atm19cold}^{WENOZ}}$. 
We find that while the disk mass is compatible with simulations including thermal effects, 
the ejecta mass is reduced. This supports our suggestion that most of the ejecta 
is caused by shock heating. 
Consequently, although we found that the ejecta mass is not affected by resolution
and is robustly around $10^{-4}\ M_\odot$, we can not rule out that the shock heating is 
artificially caused by the numerical scheme and not caused by a physical mechanism.

\section{Summary}
\label{sec:conclusions}

We have performed a detailed analysis of the conditions under which 
a supramassive neutron star can collapse to a Kerr black hole
surrounded by an accretion disk. Our approach has been two-fold:
we first analyzed the angular momentum spectrum of the collapsing configurations
and subsequently performed dynamical 3D collapse simulations to confirm our
findings.
We constructed rigidly rotating initial neutron stars using the XNS code 
\citep{Bucciantini+Del_Zanna2011, Pili+2014}. These initial 
configurations were analyzed for the mass that has enough angular 
momentum to remain outside of the ISCO of the forming black hole.
A similar study has recently been performed by \citet{Margalit:2015qza}
who used the RNS code \citep{Stergioulas:1994ea}. We argue here that, contrary to what has been
done in their work, the configuration that can collapse to a BH is 
not the maximal mass configuration, but instead the marginally stable Keplerian 
configuration, for which 
$\left. \partial M/\partial \rho_c\right|_J=0$. Moreover, we find that
a disk can form for a larger volume of the parameter space, albeit its mass is very small. Despite these
small differences we confirm their main result that it is very
difficult to form a massive disk from a collapsing neutron star
and all the cases that were investigated fall short by orders 
of magnitude to produce an energetic sGRB. These 
conclusions were subsequently confirmed by fully dynamical Numerical
Relativity simulations performed with the BAM code
\citep{Brugmann:2008zz,Thierfelder:2011yi}.

In this work, we have assumed uniform rotation and a cold EOS for the
initial configurations of the collapsing stars.
The uniform rotation is justified for the sGRB models that motivate this study.
If the supramassive NS is formed by accretion from a non-degenerate companion star \citep{macfadyen05},
there is no reason to expect differential rotation. At the moment of collapse, however,
the NS ---while being essentially cold throughout the bulk of the high-density matter---
may be engulfed by a high-temperature envelope, which is not modelled in this work.

If, in contrast, the supramassive neutron star is formed as a result of a
neutron star merger, as invoked by ``time reversal models''
\citep{Ciolfi+Siegel2015,Rezzolla:2014nva}, it is expected to be both hot ($\gtrsim10$ MeV) and
differentially rotating, at least initially. Such differentially rotating,
``hypermassive'' neutron stars can support a substantially larger mass than
rigidly rotating ones, but magnetic braking and viscosity will drive the
stars to collapse on a short time scale even if the initial seed magnetic field
is low and viscosity is small \citep{shapiro00}. Therefore, neutron stars 
that remain stable for long enough to explain the long term X-ray emission ($\sim
10^4$ s), have likely dissipated their differential rotation and have
cooled to temperatures where thermal effects in the
high-density matter are small, since the Kelvin-Helmholtz neutrino
cooling time is of the order of only seconds \citep{Radice2018}. Therefore, we consider also
in this case our assumption of essentially cold EOS and rigid rotation 
as valid.

For the equations of state expected in neutron stars ($\Gamma>2$), 
the resulting disk masses after the collapse are orders of magnitude lower 
($\lesssim 10^{-7}\ M_\odot$) than what is needed for a typical sGRB. 
Therefore, we interpret this result as disfavoring those sGRB models
that require the collapse of a supramassive NS into a BH plus disk configuration.

\section*{Acknowledgements}

  T.D.~acknowledges support by the European Union's Horizon 
  2020 research and innovation program under grant
  agreement No 749145, BNSmergers.
  Computations were performed on 
  the supercomputer SuperMUC at the LRZ
  (Munich) under the project number pr48pu
  and on the compute cluster Minerva of the 
  Max-Planck Institute for Gravitational Physics.
  
  S.R.~has been supported by the Swedish Research Council (VR) under grant
  number 2016-03657\_3, by the Swedish National Space Board under grant number
  Dnr. 107/16 and by the research environment grant  “Gravitational  Radiation
  and  Electromagnetic  Astrophysical  Transients  (GREAT)"  funded  by  the
  Swedish  Research council (VR) under Dnr. 2016-06012.

  We acknowledge support from the COST Action PHAROS (CA16214).

  We are grateful to A.~Pili, N.~Bucciantini, and L.~Del~Zanna
  for making the XNS code public available and for useful discussion.
  We thank B.~Metzger and L.~Rezzolla for their helpful comments.

\bibliographystyle{mnras}
\bibliography{refs20180807.bib}

\appendix 

\section{XNS convergence}
\label{app:XNS_vs_RNS}

\begin{figure}
\includegraphics[width=\columnwidth]{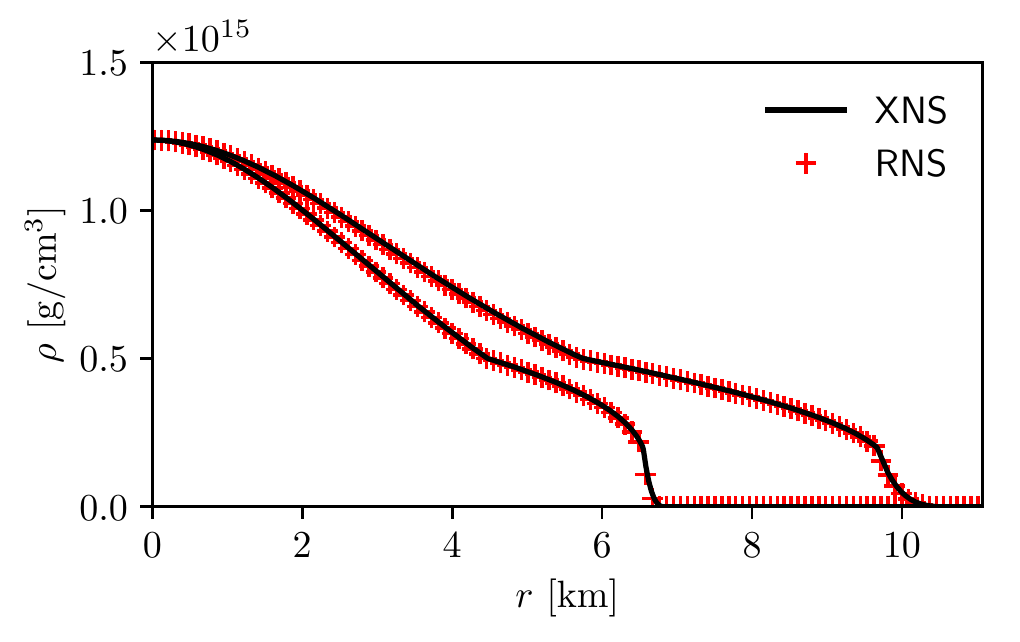}
\caption{Comparison between XNS (black lines) and RNS (red crosses) of the rest
mass density radial profiles along the equatorial and polar axes.}
\label{fig:XNS_RNS}
\end{figure}

\begin{figure*}
\includegraphics[width=0.98\columnwidth]{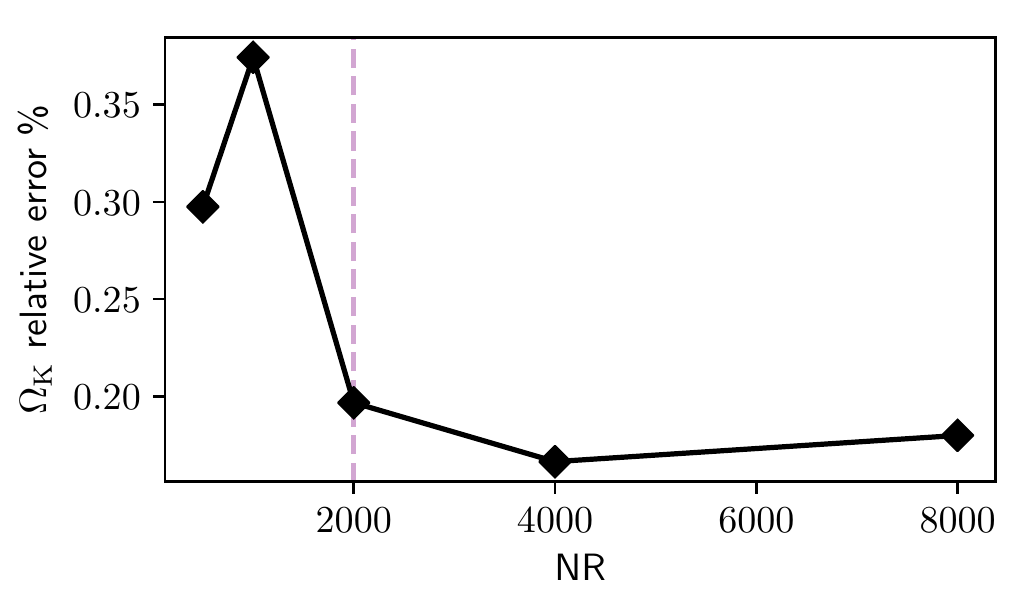}
\includegraphics[width=0.98\columnwidth]{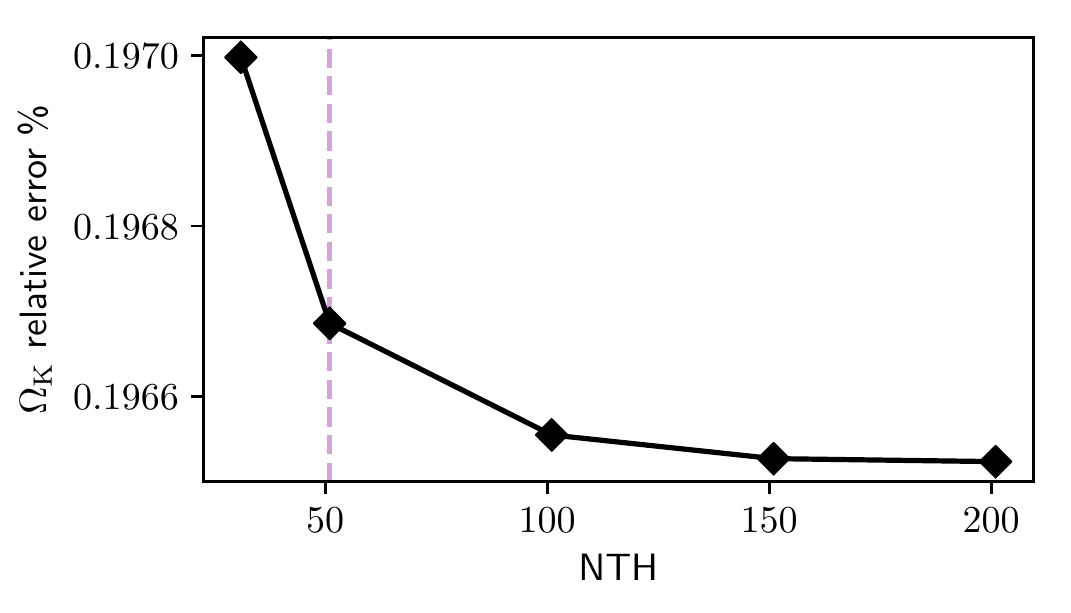}
\includegraphics[width=0.98\columnwidth]{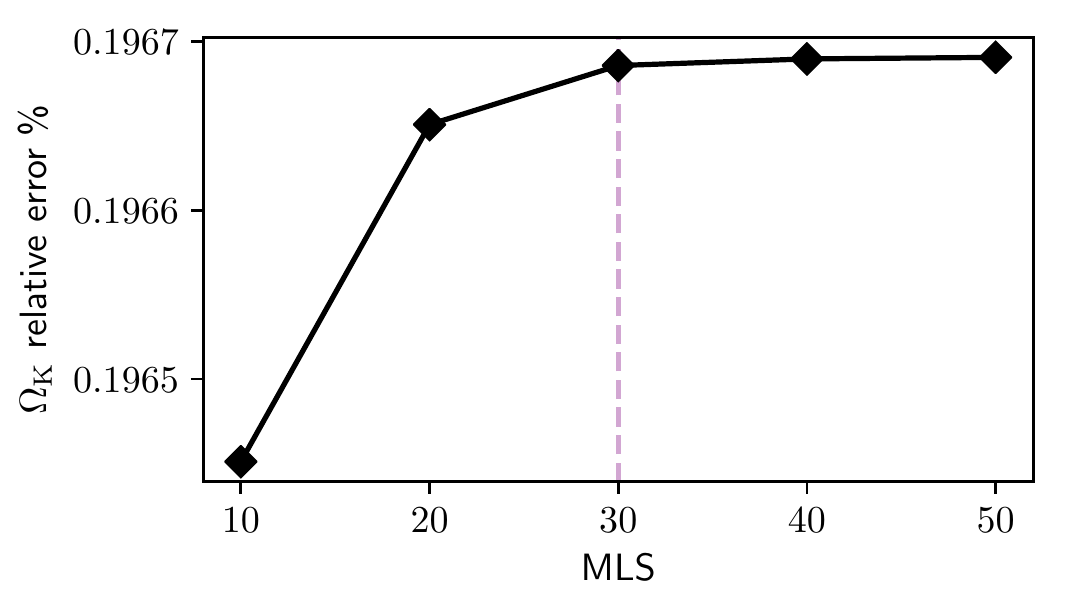}
\includegraphics[width=0.98\columnwidth]{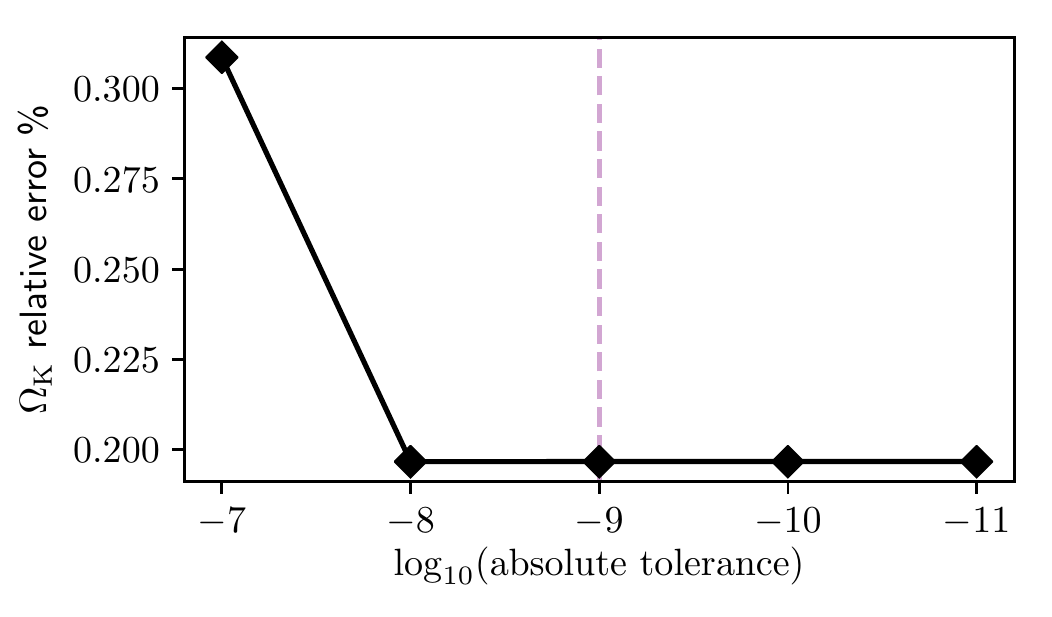}
\caption{Convergence of the Keplerian angular velocity $\Omega_{\rm K}$
obtained with XNS. We plot the relative error in percentage between the RNS
benchmark and the XNS result, varying one XNS setting every time
with respect to the baseline XNS model. From left to right and from
top to bottom, we vary respectively
the dimension of the radial grid NR, the dimension of the angular grid
NTH, the number of components in the spherical harmonics expansion MLS,
and the absolute tolerance criterion for convergence. The dashed, purple,
vertical lines mark the settings of the XNS baseline model.}
\label{fig:conv}
\end{figure*}

In this appendix we study the convergence of XNS and compare its results with the 
RNS code~\citep{RNS,Stergioulas:1994ea}.  Since we are discussing code
technical details, we will use code units, which are $c=G=M_\odot=1$.

For this purpose, we adopt the same piecewise polytrope EOS
of Case B, namely $p_1=\unit[10^{34.8}]{dyne/cm^3}$ and
$\Gamma_2=2.0$, and choose a model very close to the
marginally stable Keplerian configuration.
This is a quite demanding test,
since the EOS is not a simple 1-piece polytrope and the model
is very close to mass shedding.

\begin{table}
  \centering    
  \caption{Comparison in the model parameters in output from RNS and
  XNS. The columns contain: the stellar quantity, the RNS results,
  the XNS results, and the relative difference. The stellar quantities 
  read from top to bottom:
  the gravitational mass $M$, the total angular
  momentum $J$, the angular speed $\Omega$, the Keplerian
  angular speed $\Omega_{\rm K}$, the circumferential radius $R$,
  and the specific angular momentum at the equator $j$.}
  \begin{tabular}{cccc}
    \hline
    quantity         & RNS      & XNS      & difference\\ 
    \hline
    $M$              & 2.6394   & 2.6382   & 0.05\% \\
    $J$              & 4.6745   & 4.6373   & 0.8\% \\
    $\Omega$         & 0.040811 & 0.041000 & 0.5\% \\
    $\Omega_{\rm K}$ & 0.046103 & 0.046194 & 0.2\% \\
    $R$              & 10.668   & 10.596   & 0.7\% \\
    $j$              & 6.066    & 6.018    & 0.8\% \\
    \hline
  \end{tabular} 
 \label{tab:XNS_vs_RNS}
\end{table}

The settings of the RNS benchmark are:
\begin{itemize}
 \setlength\itemsep{-0.08em}
\item radial grid points SDIV = 3601,
\item angular grid points MDIV = 1201,
\item polynomial expansion LMAX = 36,
\item relative accuracy = $10^{-11}$,
\item EOS recovered from interpolation of a 2000 point table.
\end{itemize}
The setting of the XNS configuration, which we will call ``baseline'' model,
are the following:
\begin{itemize}
 \setlength\itemsep{-0.08em}
\item radial grid points NR = 2000,
\item angular grid points NTH = 51,
\item harmonic expansion MLS = 30,
\item absolute convergence = $10^{-9}$ (cf.~condition (iii) in Sec.~\ref{ssec:XNS}),
\item radius of the inner grid (which encompass half of the radial points) = 8,
\item radius of the outer grid = 200.
\end{itemize}
The input parameters are the central density $\rho_c=2.004\times10^{-3}$,
and the angular speed $\Omega=4.1\times10^{-2}$ for XNS and the
aspect ratio $a=0.61764727$ for RNS. We remark that
the aspect ratio in output from the baseline XNS model 
is an input for the benchmark RNS model, and should therefore
considered as ``exact'' for both codes.

In Tab.~\ref{tab:XNS_vs_RNS} we compare the output of the two codes and in
Fig.~\ref{fig:XNS_RNS} we compare the equatorial and polar density profiles.
The range of the relative differences in the model quantities between XNS and RNS
is $\sim 0.1$--$0.8\%$, except for the gravitational mass that is recovered
within $0.05\%$.
We remark that the extended conformal flatness approximation, on which
XNS is based, neglects differences between the metric functions $A$ and $B$
of the order of $\sim 0.1\%$.
Since the specific angular momentum at the ISCO of the equivalent Kerr
black hole is $j_{\rm ISCO}\simeq 7.0$ both for RNS and XNS, no debris disk is expected.
This should not surprise since the configuration is close but not equal to the
``Case B'' described in the paper, for which we expect disk formation instead.

In Fig.~\ref{fig:conv} we plot the relative differences of the Keplerian
angular velocity
between the RNS benchmark and those obtained from the XNS model with
the same settings of the XNS baseline model,
apart for the setting that is varied in each plot.
XNS shows a good convergence for each of the setting varied.

In conclusion, the two codes are in very good agreement.

\bsp	
\label{lastpage}
\end{document}